%% file: ms.tex
\documentclass[conference]{./IEEEtran}

\usepackage{cite}
\usepackage[]{graphicx}

\usepackage[utf8]{inputenc}
\usepackage[english]{babel}
\usepackage[T1]{fontenc}
\usepackage{lmodern}

\usepackage{times}

\usepackage[svgnames,x11names]{xcolor}

\usepackage{amsmath}
\usepackage{amssymb}
\usepackage{amsfonts}
\usepackage{amsfonts}
\usepackage{amsthm}
\usepackage{tabulary}
\usepackage{pifont}
\PassOptionsToPackage{hyphens}{url}
\usepackage{float}
\usepackage{adjustbox}
\usepackage{todonotes}
\usepackage{booktabs}
\usepackage{wasysym}

\input{lstconfig}

\input{circled}

\input{macros}

\usepackage{balance}

\usepackage{hyperref}

\begin{document}
\title{\toolname: Protecting Existing Smart Contracts Against Re-Entrancy Attacks}
\newcommand{\authormark}[1]{$^{#1}$}
\author{\IEEEauthorblockN{%
Michael Rodler\authormark{1}, Wenting Li\authormark{2}, %
Ghassan O. Karame\authormark{2}, Lucas Davi\authormark{1}%
}%
\IEEEauthorblockA{\authormark{1}University of Duisburg-Essen, Germany\\
\textit{\{michael.rodler,lucas.davi\}@uni-due.de}}
\IEEEauthorblockA{\authormark{2}NEC Laboratories Europe, Germany\\
\textit{wenting.li@neclab.eu}\\%
\textit{ghassan@karame.org}}
}

\IEEEoverridecommandlockouts
\makeatletter\def\@IEEEpubidpullup{6.5\baselineskip}\makeatother
\IEEEpubid{\parbox{\columnwidth}{
    To appear at\\
    Network and Distributed Systems Security (NDSS) Symposium 2019\\
    24-27 February 2019, San Diego, CA, USA\\
}
\hspace{\columnsep}\makebox[\columnwidth]{}}

\maketitle

\begin{abstract}
\input{abstract}
\end{abstract}


\section{Introduction}  
\input{sections/introduction}

\label{sec:intro}

\section{Background}
\label{sec:background}
\input{sections/background}

\section{Problem Statement and New Attacks}
\label{sec:problem}
\input{sections/problem}

\section{Design of \toolname}
\label{sec:overview}
\input{sections/overview}

\section{Implementation}
\label{sec:implementation}
\input{sections/implementation}

\subsection{Taint Tracking EVM}
\label{sec:taintrackingevm}
\input{sections/taint.tex}

\subsection{Attack Detection}
\label{sec:attackdetection}
\input{sections/reentrancy}

\section{Evaluation}
\label{sec:evaluation}
\input{sections/evaluation}

\section{Related Work}
\input{sections/relatedwork}

\section{Conclusion}
\input{sections/conclusion}

\section*{Acknowledgment}

This work has been partially funded by the DFG as part of project S2 within the CRC 1119 CROSSING.

\pagebreak

\balance
\bibliographystyle{./IEEEtranS}
\bibliography{IEEEabrv,references}

\end{document}

%% file: lstconfig.tex
\definecolor{lstgreen}{rgb}{0,0.7,0}
\usepackage{listings}
\lstdefinelanguage{solidity}{
  morekeywords={new, true, false, function, return,
  switch, var, if, in, while, do, else, case, break, 
  contract, mapping, struct, library, 
  revert, assert, require, throw, sha3, keccak256,
  sstore, sload, delete},
  morekeywords=[2]{export, public, private, payable, view, returns},
  morekeywords=[3]{bool, uint, int, address, uint256, int256, int8, uint8,
    uint128, int128, bytes, bytes32, bytes4},
  morekeywords=[4]{msg, this, tx},
  sensitive=false,
  comment=[l]{//},
  morecomment=[s]{/*}{*/},
  morestring=[b]',
  morestring=[b]"
}

\definecolor{lstgreen}{rgb}{0,0.6,0}
\usepackage{listings}
\lstdefinestyle{plain}{
  numbers=none,
  frame=none,
  xleftmargin=1pt,
  xrightmargin=1pt,
}

%% file: circled.tex
\usepackage{tikz}
\usetikzlibrary{shapes,arrows,positioning,shapes.multipart}

\newcommand*\circleds[1]{\tikz[baseline=(char.base)]{%
            \node[shape=circle,draw,inner sep=1pt] (char) {\tiny #1};}}

\newcommand{\circleone}{\ding{192}}
\newcommand{\circletwo}{\ding{193}}
\newcommand{\circlethree}{\ding{194}}
\newcommand{\circlefour}{\ding{195}}

%% file: macros.tex
\newcommand{\evm}{EVM}

\newcommand{\theDAO}{``TheDAO''}

\newcommand{\toolname}{\emph{Sereum}}

\newcommand{\victim}{\emph{Victim}}
\newcommand{\attacker}{\emph{Attacker}}

\newcommand{\eInst}[1]{\texttt{#1}}
\newcommand{\eVar}[1]{\textit{#1}}
\newcommand{\eFunc}[1]{\textit{#1}}
\newcommand{\funcname}[1]{\textit{#1}}

\renewcommand{\paragraph}[1]{\noindent\textbf{#1}}

\makeatletter
\newcommand*{\rom}[1]{\expandafter\@slowromancap\romannumeral #1@}
\makeatother

%% file: abstract.tex
Recently, a number of existing blockchain systems have witnessed major bugs and
vulnerabilities within smart contracts. Although the literature features a
number of proposals for securing smart contracts, these proposals mostly focus
on proving the correctness or absence of a certain type of vulnerability within
a contract, but cannot protect deployed (legacy) contracts from being
exploited.
In this paper, we address this problem in the context of re-entrancy exploits
and propose a novel smart contract security technology, dubbed \toolname\
(Secure Ethereum), which protects existing, deployed contracts against
re-entrancy attacks in a backwards compatible way based on run-time monitoring
and validation. \toolname\ does neither require any modification nor any
semantic knowledge of existing contracts. By means of implementation and
evaluation using the Ethereum blockchain, we show that \toolname\ covers the
actual execution flow of a smart contract to accurately detect and prevent
attacks with a false positive rate as small as 0.06\% and with negligible
run-time overhead. As a by-product, we develop three advanced re-entrancy
attacks to demonstrate the limitations of existing offline vulnerability
analysis tools. 

%% file: sections/introduction.tex

The massive adoption of Bitcoin has fueled innovation, and there are
currently more than 500 alternative blockchains---most of which are simple
variants of Bitcoin~\cite{altcoins-list}. Bitcoin unveiled a key-enabling technology and a hidden
potential, the \emph{blockchain}. Indeed, the blockchain
allows transactions, and any other data, to be securely stored and verified
without the need of any centralized authority.
Currently, a number of blockchains, such as Ethereum, provide means to execute
programs on the blockchain. These programs are referred to as smart contracts
and allow nearly arbitrary (Turing-complete) business logic to be implemented.
In Ethereum, smart contracts are, besides the Ether cryptocurrency, a crucial
part of the blockchain.
Ethereum allows to attach a smart
contract program to an address. When a transaction involves such an address,
the nodes in the Ethereum network will execute the contract, which can trigger
further transactions, update state on the blockchain, or simply abort the
transaction.

\vfill\null

In blockchain systems, such as Ethereum, smart contracts are capable of owning
and autonomously transferring currency to other parties.  As such, it is vital
that smart contracts execute correctly and satisfy the intention of all
stakeholders. Recently, the blockchain community has witnessed a number of
major bugs and vulnerabilities within smart contracts.  In some cases,
vulnerabilities allowed an attacker to maliciously extract currency from a
contract. For instance, the infamous attack on the \theDAO\ smart contract
resulted in a loss of over 50 million US Dollars worth of Ether at the time the
attack occurred~\cite{daohackamount}. The DAO attack is an instance of a
re-entrancy attack where the main contract calls an external contract which
again calls into the calling contract within the same transaction.


These attacks have fueled interest in the community to conduct research on
solutions dedicated to enhance the security of smart contracts. Recently presented
approaches range from devising better development environments to using safer programming
languages~\cite{Coblenz2017-obsidian}, formal verification \cite{zeus-ndss2018}
and symbolic execution \cite{luu2016making}. Prior work has focused primarily
on techniques that detect and prevent possible vulnerabilities upfront.
For instance, Oyente~\cite{luu2016making} proposed using symbolic execution to
find vulnerabilities in smart contracts. ZEUS~\cite{zeus-ndss2018} uses model
checking to verify the correctness of smart contracts, and Securify~\cite{securifych}
performs advanced static analysis to infer semantic facts
about data-flows in a smart contracts to prove the presence or absence of
vulnerabilities.
Other recent approaches use symbolic execution to automatically construct
exploits in order to demonstrate the vulnerability of an analyzed smart
contract~\cite{teether,maian}.

\noindent
\textbf{Challenges in Fixing Smart Contracts.}
We note that fixing discovered bugs in smart contracts is particularly challenging due to three key challenges: \emph{(1)~the code of a smart contract is
expected to be immutable after deployment, (2)~smart contract owners are anonymous, i.e.,
responsible disclosure is usually infeasible, and (3)~existing approaches are mostly performing offline analysis
and are susceptible to missing unknown run-time attack patterns}. As a consequence of (1), approaches that
prove correctness or absence of a certain type of
vulnerability~\cite{luu2016making,securifych,zeus-ndss2018} are only important
for the development of future smart contracts, but leave already deployed
(legacy) contracts vulnerable. More specifically, to deal with a
vulnerable contract and restore a safe state, the owner
of the contract must deprecate the vulnerable contract, move all funds out of
the contract, deploy a new contract, and move the funds to
the new contract. This process is largely cumbersome since
the address of the vulnerable contract might be referenced by other contracts
(see for example~\cite{eth-upgrade-strategies}).
Even if this process could be simplified, it remains still unclear how to contact
contract owners to inform them about contract vulnerabilities. For instance, a recent study was
able to generate exploits to steal Ether from 815~existing smart contracts. However, the authors
refrained from mentioning any particular smart contract as it was not possible to
report the discovered bugs to \emph{any} of the creators~\cite{teether}.
Finally, offline analysis techniques typically cannot fully cover the run-time behavior of a
smart contract thereby missing novel attacks exploiting
code constructs that were believed to be not exploitable.

\noindent
\textbf{Research Question.}
Given these challenges, \emph{this paper aims to answer the question whether we
can protect legacy, vulnerable smart contracts from being exploited without (1)~changing the smart contract code, and (2)~possessing any semantic knowledge on the smart contract}. To answer
this question, we focus our analysis on re-entrancy attacks. Among the
attack techniques proposed against smart contracts~\cite{atzei2017survey},
re-entrancy
attacks play a particular role as they have been leveraged in the DAO attack~\cite{daohackamount} which is undoubtedly the most popular smart contract attack until today. Recent studies also argue that many smart contracts are vulnerable to re-entrancy, e.g., Oyente reports 185 and Securify around 1,400 contracts as vulnerable to re-entrancy attacks~\cite{luu2016making,securifych}. In addition, re-entrancy attack patterns are suitable for run-time detection given the conditions mentioned in our
research question (i.e., no code changes, no prior knowledge).
Surprisingly, as our
systematic investigation reveals, new classes of re-entrancy attacks, beyond DAO, can be developed \emph{without being detected} by the plethora of existing defenses proposed in the literature, such as~\cite{luu2016making,securifych,zeus-ndss2018}.

\noindent
\textbf{Contributions.}
In this paper, we present the design and implementation of a novel smart contract
security technology, called \toolname\ (Secure Ethereum), which is able to protect \emph{existing, deployed contracts}
against re-entrancy attacks in a \emph{backwards compatible} way by performing \emph{run-time monitoring} of
smart contract execution with negligible overhead. Given our run-time monitoring technique, \toolname\ is able to cover
the actual execution flow of a smart contract to accurately detect and prevent attacks.
As such, our approach also sheds important lights on the general problem of incompleteness of any offline,
static analysis tool. To underline this fact, \emph{we develop three new re-entrancy attacks} in Section~\ref{sec:problem}
(cross-function, delegated, and create-based re-entrancy) that undermine
existing vulnerability detection tools~\cite{luu2016making,securifych} but are
detected in \toolname.

Our prototype implementation (cf. Section~\ref{sec:implementation}) targets the Ethereum Virtual Machine (EVM)
which is currently the most popular platform for running smart contracts. In this context, we introduce a hardened EVM
which leverages taint tracking to monitor execution of smart contracts.
While taint tracking is a well-known technique to detect leakage of private
data~\cite{enck2014-taintdroid} or memory corruption attacks~\cite{Clause2007-dytan}, we
apply it for the first time to a smart contract execution platform. Specifically,
we exploit taint analysis to monitor data flows from storage variables to control-flow decisions.
Our main idea (cf. Section~\ref{sec:overview}) is to introduce write locks, which prevent the contract from updating storage
variables in other invocations of the same contract of one Ethereum
transaction. \toolname\ prevents any write to variables, which would render
the contract's state inconsistent with a different re-entered execution of the
same contract. \toolname\ also rolls back transactions that trigger an invalid write
to variables---thereby effectively preventing re-entrancy attacks.
\toolname\ can also be used as a passive detection tool, where it does not
rollback attack transactions, but only issues a warning for detected attacks.
%

We perform an extensive evaluation of our \toolname\ prototype by re-executing
a large subset of transactions of the Ethereum blockchain (cf. Section~\ref{sec:evaluation}). Our results show that \toolname\ detects all malicious transactions related to
the DAO attack, and only incurs $9.6\%$ run-time overhead; we further verify our findings by using existing vulnerability detection tools
and manual code analysis on selected contracts. Although \toolname\ only results
in 0.06\%~of false positives, we provide a thorough investigation of
false positive associated with our approach and other existing static analysis tools~\cite{luu2016making,securifych} thereby demonstrating
that \toolname\ provides improved detection of re-entrancy attacks compared to existing approaches with negligible run-time overhead.

%% file: sections/background.tex

In this section, we recall the basics of smart contracts and the Ethereum
Virtual Machine (Section~\ref{sec:evm}). We also describe the
implementation details of existing re-entrancy attacks
(Section~\ref{sec:re-entrancy}), and discuss common defense techniques
against these attacks (Section~\ref{sec:defenses}).

\subsection{Smart Contracts and the Ethereum Virtual Machine}\label{sec:evm}

In general, the blockchain consists of a distributed ledger where transactions
are committed in the same order across all nodes. 
Smart contracts typically consist of self-contained code that is executed by
all blockchain nodes. The execution of smart contracts is typically confined
to a deterministic context (e.g., based on the same input, ledger state,
run-time environment) which is replicated on benign nodes. This
ensures that the state update on the ledger is propagated to all nodes in
the network.
%

The currently most popular blockchain technology for smart contracts is Ethereum~\cite{Wood2016-yellowpaper}.
Ethereum smart contracts receive and send the cryptocurrency Ether.
Contracts are invoked through transactions which are issued either by Ethereum
clients or other contracts.
Transactions need to specify the invoked contract functions, which are public interfaces exposed by the
target contracts.
In order to incentivize the network to execute contracts, Ethereum
relies on the mechanism of \emph{gas}: the amount of gas corresponding to a contract relates to
the cost of executing that contract and is paid along with the invocation transaction by the
sender in Ether to fuel the execution of a contract.
This mechanism also prevents vulnerable code (e.g., infinite loops) from harming the entire network.
%
%

Although Ethereum supports several programming languages and compilers, the most common
language for Ethereum contracts is currently Solidity~\cite{solidity}.
The bytecode of contracts (generated by the Solidity compiler \emph{solc}) is
distributed via dedicated contract creation transactions and gets executed by the
\evm\ on each local node. Once the contract creation
transaction is committed to the ledger,
all nodes compute the \emph{contract address}---which is required to invoke
contracts---and initialize the contract code and data.

\noindent
\textbf{Ethereum Virtual Machine (EVM).}
The \evm\ follows the stack machine architecture, where
instructions either pop operands from the data stack or use constant operands.
The overall architecture of the \evm\ is tailored towards the peculiarities
of blockchain environments~\cite{Wood2016-yellowpaper}:
\begin{itemize}
	\item \textbf{Execution Context}:
	To ensure that a transaction execution is deterministically,
	all environmental information is fixed with respect to the block where the
	transaction is contained.
	For instance, a contract cannot use the system time. Instead, it must use the current block
	number and timestamp.
	\item \textbf{Memory}:
	An \evm\ contract can use three different memory regions to store
	mutable data during the execution: stack, memory and storage.
	The \emph{stack} is a volatile memory region whose content can only be changed with dedicated instructions.
	The \evm\ distinguishes the \emph{call stack} (maximum depth 1024) from the \emph{data
	stack}. The so-called \emph{memory} is a volatile heap-like memory region, where every byte is addressable.
	The only state persistent across transactions is maintained in the \emph{storage} region
	which can be thought of a key-value store that maps 256-bit words to 256-bit
	words.
	\item \textbf{Procedure Calls}:
	The \evm\ \texttt{CALL} instruction can be considered as a
	Remote Procedure Call (RPC) as it transfers control to
	another (untrusted) contract. \texttt{DELEGATECALL} is
	similar to \texttt{CALL} with the difference that the
	invoked contract shares the same execution context as the caller.
	Consecutive calls are pushed to the \evm\ call stack; an exception will be thrown
	once the maximum call stack depth is reached.
\end{itemize}

\subsection{Re-entrancy Problem} \label{sec:re-entrancy}



\begin{figure}[t]
  \centering
  \input{figures/withdraw-listing}
  \includegraphics[width=0.9\linewidth]{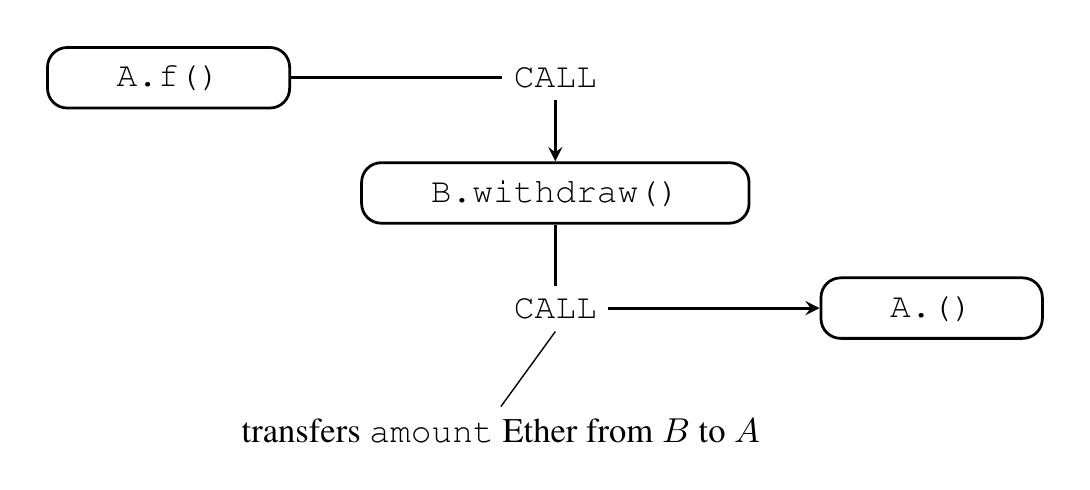}
  \caption{Common withdrawal pattern in Solidity: the upper part shows the
sample Solidity code, whereas the lower part shows the call chain. In this
example contract $A$ withdraws 100~wei from contract $B$.}
  \label{fig:withdraw}
\end{figure}

Re-entrancy attacks emerge as one of the most severe and effective attack vectors against smart contracts.
Re-entrancy of a contract occurs when a contract calls another (external) contract which again calls
back into the calling contract. All these actions are executed within a single transaction.
Legitimate re-entrancy often happens during normal contract execution, as it is
part of common and officially supported programming patterns for Ethereum smart
contracts~\cite{solidity-withdraw}.
Consider the common withdrawal pattern~\cite{solidity-withdraw} depicted in Figure~\ref{fig:withdraw} which
shows how contract~$A$ withdraws $100$~wei from contract~$B$. The key rationale
of re-entrancy is to allow other contracts to withdraw funds from their balance.
In Figure~\ref{fig:withdraw}, contract $A$ invokes the public \emph{withdraw} function
of contract $B$, whereas $B$ subsequently invokes the \emph{msg.sender.send.value} function
to transfer the specified amount to $A$ (i.e., \emph{msg.sender} is representing the calling contract $A$).
In Ethereum, Ether is transferred by means of a function
call, e.g., contract $B$ must call back (\emph{re-enter}) into contract $A$'s
fallback function
to send the funds.
The fallback function is indicated by the function without function name.

To support calling other contracts Solidity supports two high-level
constructs for calling into another contract: \eFunc{send} and \eFunc{call}.
Both are implemented as \eInst{CALL} instructions on the \evm\ level.
However, if the recipient account is another contract, \eFunc{send} only invokes the
fallback function of the recipient contract, while \eFunc{call} allows the
caller to specify any function signature of the recipient contract.
Further, \eFunc{send} only supplies a limited amount of
gas. The limited amount of gas,
which is provided by \eFunc{send}, prevents the called contract from performing
other gas-expensive instructions, such as performing further calls.
While re-entrancy is necessary for the withdrawal pattern and several other
programming patterns~\cite{solidity-withdraw}, it can be exploited if not carefully implemented, e.g.,
loss of 50~million US Dollars in the case of DAO~\cite{dao-attack,daohackamount}.

\begin{figure}[t]
  \centering
  \input{figures/simpledao-reenter-listing}
  \includegraphics[width=\linewidth]{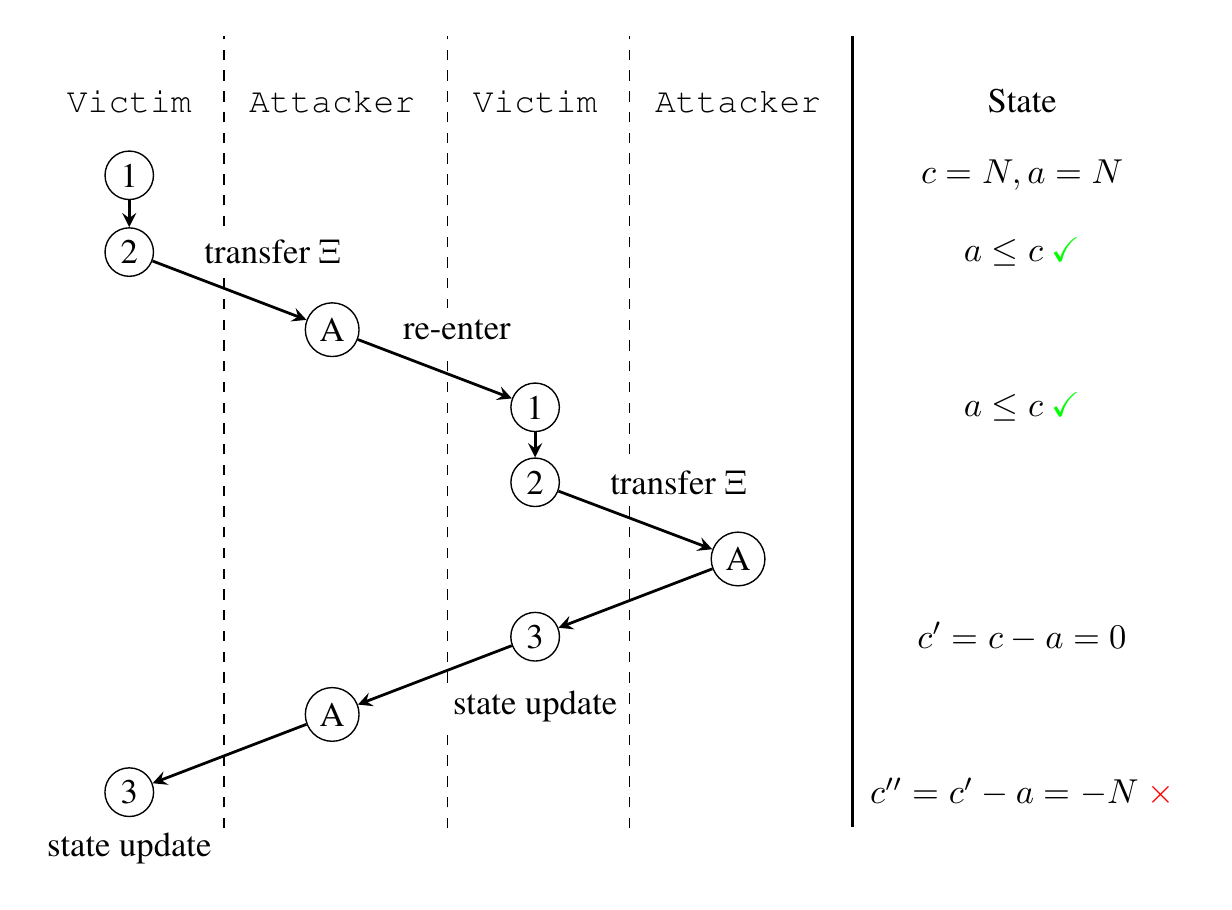}
  \caption{Sample contract vulnerable to  re-entrancy attacks~\cite{atzei2017survey}: the upper parts shows the Solidity code, whereas the lower part shows the call sequence between the vulnerable contract \texttt{Victim} and the attacker contract, and the state of the variable $a$ (amount) and $c$ (credit[msg.sender]). The amount $a$ has not been updated for the second invocation of \texttt{Victim} thereby allowing a malicious re-entrancy.}
  \label{fig:simpledaowithdraw}
\end{figure}

A malicious re-entrancy occurs when a contract is re-entered \emph{unexpectedly}
and the contract operates on \emph{inconsistent} internal state.
More specifically, if a re-entrance call involves
a control-flow decision that is based on some internal state of the victim
contract, and the state is updated \emph{after} the external call returns, then
it implies that the re-entered victim contract operated based on an inconsistent
state value, and thus the re-entrancy was not expected by the contract developer.
For example, Figure~\ref{fig:simpledaowithdraw} shows a simplified version of
a contract (inspired by~\cite{atzei2017survey}), called \victim, which suffers from a re-entrancy
vulnerability. \victim\ keeps track of an amount ($a$) and features
the \eFunc{withdraw} function allowing other contracts to withdraw Ether ($c$).
The \eFunc{withdraw} function must perform three steps: \circleone\ check whether
the calling contract is allowed to withdraw the
requested amount of Ether, e.g., checking whether $a \le c$, \circletwo\ send the amount of Ether to the calling contract
and \circlethree\ update the internal state to reflect the new amount, e.g., $c-a$.
Note that step \circletwo\ is performed before the state is updated in
\circlethree. Hence,
a malicious contract, can re-enter the contract and call
\texttt{withdraw} based on the same conditions and amounts as for the first invocation.
As such, an attacker can repeatedly re-enter into \victim\ to transfer large amounts of Ether
until the \victim\ is drained of Ether.
A secure version of our simple example requires swapping
lines 3 and 4 to ensure that the second invocation of \victim\ operates on
consistent state with updated amounts. In Section~\ref{sec:problem}, we elaborate
on the challenges of fixing vulnerable contracts and the prevalence of re-entrancy vulnerabilities in
existing contracts.

\subsection{Common Defenses and Analysis Tools}\label{sec:defenses}

While this paper puts its focus on re-entrancy vulnerabilities, other types of
vulnerabilities have also been discovered (e.g., integer overflow, type
confusion) that have been comprehensively surveyed in~\cite{atzei2017survey}.
To combat smart contract vulnerabilities, the literature features a number of
proposals and tools for identifying vulnerabilities in smart contracts.
For instance, \emph{Oyente}~\cite{luu2016making}, \emph{Mythril}~\cite{mythril}
and \emph{Manticore}~\cite{manticore} leverage symbolic
execution~\cite{king1976symbolic} to detect various types of bugs (including
re-entrancy) in Ethereum smart contracts.
\emph{teEther}~\cite{teether} is a tool that automatically generates exploits
for smart contracts. It defines the notion of vulnerable state, in which Ether
can be transferred to an attacker-controlled address. By means of symbolic
execution, a transaction sequence can be inferred to reach the vulnerable
state. This transaction sequence is used to automatically generate the exploit.
Similarly, \emph{Maian}~\cite{maian} relies on symbolic analysis, but aims at
finding a sequence of invocations that construct traces that lead to
vulnerabilities.
However, symbolic execution techniques suffer from the well-known \emph{path
explosion} problem for larger programs which is still an ongoing research
topic~\cite{sok-taint-analysis,Kuznetsov2012-gp,Avgerinos2014-fw,Trabish2018-ku}.

\emph{Zeus}~\cite{zeus-ndss2018} introduces a policy language to assert the
correctness as well as the security requirements of a contract. For this, it
requires contract source code and user-defined policies. It applies static
analysis based on symbolic verification to find assertion violations.
\emph{SmartCheck}~\cite{smartcheck} first
converts the solidity contract source code to a XML-based parse-tree and then
searches for vulnerable patterns through XPath queries.
\emph{Securify}~\cite{securifych} uses static analysis to infer
semantic facts about smart contracts. These semantic facts are passed to a
Datalog solver~\cite{datalog}, which can prove whether a
defined compliance pattern or violation pattern is satisfied thereby proving
the absence or presence of certain vulnerabilities.
Other works leverage translation to F* to prove safety and security properties
of smart contracts and to improve on existing static analysis
tools~\cite{maffei-staticanalysis,fstar-verify}.
\emph{KEVM}~\cite{hildenbrandt2017kevm} defines executable formal semantics for
\evm\ bytecode in the $\mathbb{K}$-framework and presents an accompanying
formal verification tool.

\emph{ECFChecker}~\cite{ecfchecker} is an analysis tool that detects re-entrancy
vulnerabilities by defining a new attribute, Effectively Callback
Free (ECF). An execution is ECF when there exists an equivalent execution without callbacks that can achieve the same
state transition. If all possible executions of a contract satisfy ECF, the
whole contract is considered as featuring ECF.
Non-ECF contracts are thus considered as
vulnerable to re-entrancy, as callbacks can affect the state transition upon
contract execution.
Proving the ECF property statically was shown to be undecidable in general.
However, Grossman et al. also developed a dynamic checker that can show whether
a transaction violates the ECF property of a contract~\cite{ecfchecker}.
\emph{ECFChecker} has been developed concurrently to \toolname\ and is, to the
best of our knowledge, the only other runtime monitoring tool. However, as we
argue in Section~\ref{sec:problem}, this approach does not cover the full space
of re-entrancy attacks.


%% file: figures/withdraw-listing.tex
\begin{lstlisting}[language=solidity]
contract A {
  function f() { b.withdraw(100); }
  function () public payable { }
}

contract B {
  function withdraw(uint amount) public {
    msg.sender.send.value(amount)();
  }
}
\end{lstlisting}

%% file: figures/simpledao-reenter-listing.tex
\begin{lstlisting}[escapechar=!]
function withdraw(uint amount) public {
 !\circleds{1}! if (credit[msg.sender] >= amount) { 
 !\circleds{2}!   msg.sender.call.value(amount)(); 
 !\circleds{3}!   credit[msg.sender] -= amount; 
  }
}
\end{lstlisting}

%% file: sections/problem.tex

In this paper, we set out to propose a defense (cf. Section~\ref{sec:overview})
which protects existing, deployed smart contracts against re-entrancy
attacks in a backwards-compatible way without requiring source code
or any modification of the contract code. As mentioned earlier, re-entrancy patterns are
prevalent in smart contracts and require developers to carefully follow
the implementation guidelines~\cite{solidity-withdraw}.

As the attack against
\theDAO\ demonstrated, contracts that are vulnerable re-entrancy attacks can be
drained of all Ether.
Until now, the only publicly documented re-entrancy attack, was against the \theDAO\
contract~\cite{daohackamount}. Our evaluation also shows that
re-entrancy attacks have not \emph{yet} been launched against other contracts
(except some new minor incidents we will describe in Section~\ref{sec:evaluation}).
However, recent studies demonstrate that many already deployed contracts are
vulnerable, e.g., Oyente flags 185~
contracts as potentially vulnerable. These findings demonstrate that a
systematic defense against re-entrancy attacks is urgently required to protect
these contracts from being exploited.

As discussed in Section~\ref{sec:defenses}, the majority of defenses
deploy static analysis and symbolic execution techniques to
identify re-entrancy vulnerabilities. While these tools surely help
in avoiding re-entrancy for new contracts, it remains open how to protect
existing contracts. That is, fixing smart contract vulnerabilities based on these tools
is highly challenging owing to the immutability of smart contract
code and anonymity of smart contract owners (cf. Section~\ref{sec:intro}).

Apart from these
fundamental limitations, we also observe that existing approaches fail
to effectively detect all re-entrancy vulnerabilities or suffer from a high
number of false positives.
More specifically, we note that existing
approaches can be undermined by advanced re-entrancy attacks.
%
To this end, we identify three re-entrancy patterns, which existing tools do
not flag as re-entrancy vulnerabilities but are nevertheless exploitable.
We call these patterns (1)~\emph{cross-function re-entrancy},
(2)~\emph{delegated re-entrancy} and (3)~\emph{create-based re-entrancy}.
While cross-function re-entrancy vulnerabilities
have been partially discussed in the Ethereum community
(e.g.,~\cite{solidity-cross-re-enter-2,solidity-cross-re-enter}),
we believe that this is first presentation of delegated and create-based re-entrancy attacks.
All of these attacks are either missed or imprecisely detected by the state-of-the-art
detection tools such as Oyente~\cite{luu2016making}, Securify~\cite{securifych}, and ZEUS~\cite{zeus-ndss2018}.

In what follows, we present three attacks that exploit these re-entrancy
patterns and discuss why existing tools cannot accurately mark the contract
code as vulnerable. As we show, these attacks map to standard programming
patterns and are highly likely to be included in existing contracts.
%
For the purpose of re-producing our attacks and testing them against the public
detection tools, the source codes of the vulnerable contracts and the
corresponding attacks is available at~\cite{ude-projectpage}.

\input{sections/attacks}

%% file: sections/attacks.tex

\subsection{Cross-Function Re-Entrancy}
\label{sec:attackcrossfunction}
The first attack that we developed exploits the fact that a re-entrancy attack
spans over multiple functions of the victim contract. We show that such
cross-function re-entrancy attacks are equally dangerous as traditional
same-function re-entrancy. In classical re-entrancy attacks the same function
of the contract is re-entered again. In cross-function re-entrancy the same
contract is re-entered in a different function.
This attack exploits the fact that smart contracts often offer multiple
interfaces, that read or write the same internal state variables.

\begin{figure}[t]
  \centering
  \input{figures/cross-func-listings}
  \includegraphics[width=0.9\linewidth]{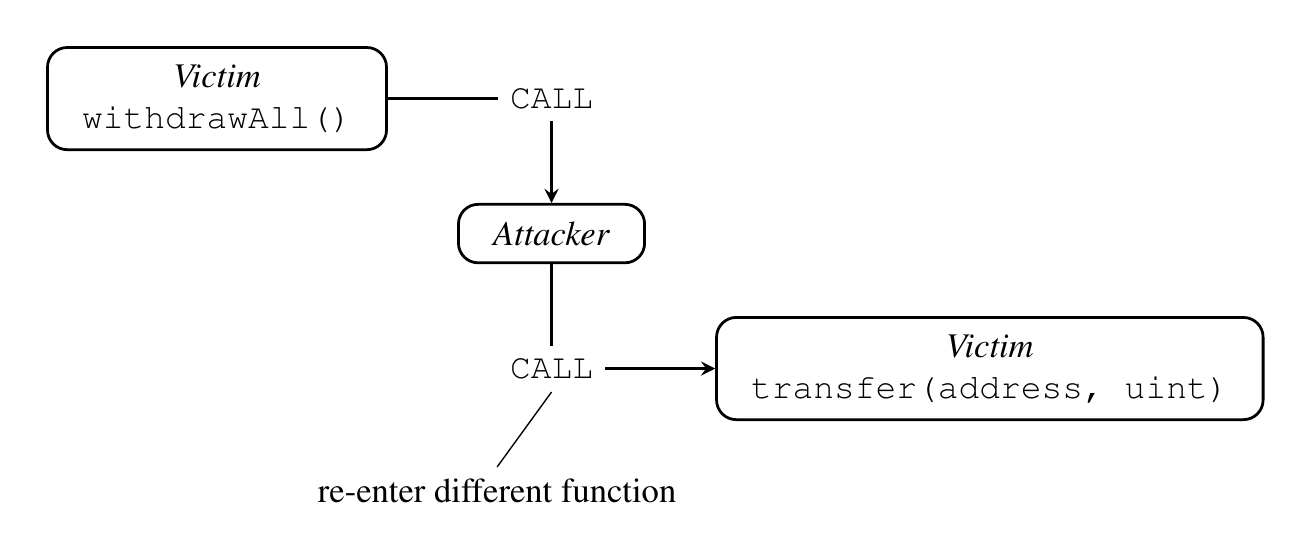}
  \caption{The upper part shows the relevant code for a customized ERC20 Token
    with a cross-function re-entrancy bug. The lower part shows the call chain
    during the attack. The attacker cannot re-enter \emph{withdrawAll}.
    However, the \emph{transferToken} can still be re-entered and abused to
    transfer tokens to another attacker-controlled address. We assume the
  attacker is then able to exchange the tokens for Ether.}
  \label{fig:crossfuncreenter}
\end{figure}

For the sake of an example, consider the snippet from an ERC20 Token like contract depicted in
Figure~\ref{fig:crossfuncreenter}. The function \eFunc{withdrawAll} performs a
state update (the update of \eVar{tokenBalance}) after an external call.
However, an attacker cannot simply re-enter the \eFunc{withdrawAll} function since
the \eVar{etherAmount} is set to zero before the external call.
Thus, the condition check in line 7 cannot evaluate to
true anymore thereby preventing re-entrancy. However, the attacker can still trigger re-entrancy on other
functions. For instance, the attacker can re-enter the \eFunc{transfer}
function, which uses the inconsistent \eVar{tokenBalance} variable. This allows
the attacker to transfer tokens to another address, although the attacker
should not have any token available anymore.

Unfortunately, existing academic static analysis tools do not accurately address cross-function re-entrancy.
Namely, Oyente does not flag the code depicted in
Figure~\ref{fig:crossfuncreenter} as vulnerable to re-entrancy. Securify and
Mythril apply a
too conservative policy with regards to re-entrancy in general: both flag any
state update occurring after an external call as a bug not considering whether
the state update actually causes inconsistent state. Hence, it suffers from
significant false positive issues that we will discuss in more detail in Section~\ref{sec:evaluation}.

In general, detecting cross-function re-entrancy is challenging for any static analysis
tool due to the potential state explosion in case every external
call is checked to be safe for every function of the contract. For exactly this reason,
ZEUS omitted to perform any cross-function analysis~\cite{zeus-ndss2018}.
However, recent work in symbolic execution tools allows detection of
cross-function re-entrancy vulnerabilities. For example,
Manticore~\cite{manticore} is able to detect cross-function re-entrancy
attacks.
In general, ECFChecker is able to detect cross-function re-entrancy attacks.
However, during our evaluation, we were able to construct a contract that can
be exploited with a cross-function re-entrancy attack without being detected by
ECFChecker. We include this specific contract as part of our set of vulnerable
contracts~\cite{ude-projectpage}.

\subsection{Delegated Re-Entrancy}
\label{sec:attackdelegated}

Our second attack performs a new form of re-entrancy that hides the
vulnerability within a \eInst{DELETEGATECALL} or \eInst{CALLCODE} instruction. These EVM instructions
allow a contract to invoke code of another contract in the context of
the calling contract. These instructions are mostly used to implement dynamic library
contracts. In Ethereum libraries are simply other contracts deployed on the blockchain.
When a contract invokes a library, they share the same execution context. A
library has full control over the calling contracts funds and internal
state, i.e., the storage memory region.
Using libraries has the advantage that many contracts can re-use the same
code, which is deployed only once on the blockchain. Furthermore, it
also allows a contract to update functionality by switching to a newer
version of the library.

\begin{figure}[t]
  \centering
  \input{figures/delegated-listing}
  \includegraphics[width=0.6\linewidth]{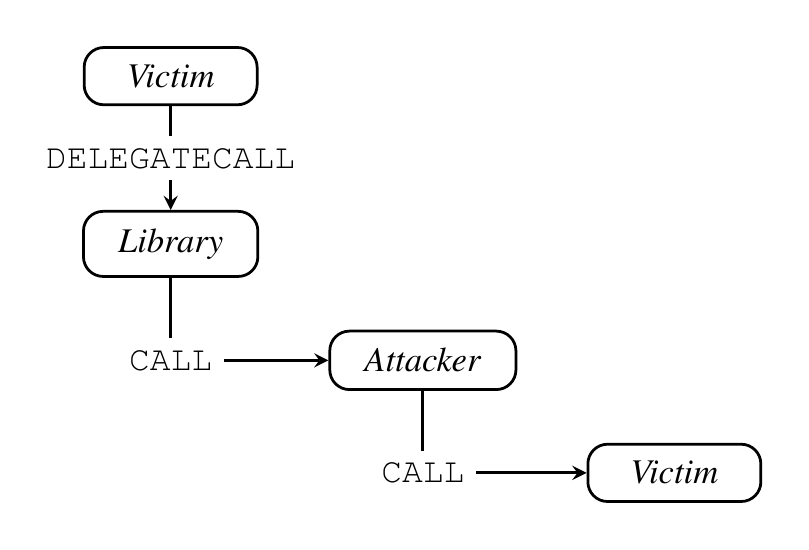}
  \caption{The upper part show the relevant solidity source code. The lower
    part shows the call chain for a delegated re-entrancy attack. Analyzed in
    isolation, the \victim\ and the \textit{Library} contract are not vulnerable
    to re-entrancy. However, when the \victim\ contract is combined with the
  \textit{Library} contract, it becomes vulnerable. In this simplified case the
\textit{Library} contract is simply used for sending Ether.}
  \label{fig:delegated}
\end{figure}

For a combination of contract and libraries to be vulnerable, the state-update
and the external call must take place in different contracts. For example, the
improper state-update happens in one library after the contract already
performed the external call. When each one is analyzed in isolation, none of
the contracts exhibit a re-entrancy vulnerability. However, when both contracts
are combined, a new re-entrancy vulnerability emerges which we refer to as
delegated re-entrancy.
Figure~\ref{fig:delegated} shows a simplified example of a contract, which uses
a library contract for issuing external calls.

Existing static analysis tools cannot detect delegated re-entrancy attacks:
during offline analysis, it is not known which library contract will be used when
actually executing the smart contract. Hence, existing
analysis tools, such as Oyente or Securify, fail to identify the delegated
re-entrancy vulnerability as they analyze contracts in isolation.
Although symbolic execution techniques could potentially leverage the current blockchain
state to infer which library is eventually called and dynamically fetch the code
of the library contract, this is not a viable solution as a
future (updated) version of the library might introduce a new vulnerability. To
detect these attacks, a run-time solution emerges as one of the few workable
and effective means to deter this attack.
Due to its dynamic nature, ECFChecker is able to detect delegated re-entrancy
attacks, as it analyzes the actual combination of contracts and libraries.

\subsection{Create-Based Re-Entrancy}
\label{sec:attackcreatebased}

Similar to delegated re-entrancy attacks, our third type of attack exploits the
fact that a contract's constructor can
issue further external calls. Recall that contracts can either be created by
accounts (with a special transaction) or by other contracts.
In solidity, a new contract can be created with the \emph{new} keyword. On the
\evm\ level, this is implemented with the \eInst{CREATE} instruction. Whenever a
new contract is created, the constructor of that contract will be executed
immediately. Usually, the newly created contract will be trusted and as such
does not pose a threat. However, the newly created contract can issue further
calls in its constructor to other, possibly malicious, contracts.
To be vulnerable to a create-based re-entrancy attack, the victim contract must
first create a new contract and afterwards update its own internal state,
resulting in a possible inconsistent state. The newly created contract must
also issue an external call to an attacker-controlled address. This then allows
the attacker to re-enter the victim contract and exploit the inconsistent
state.

Create-based re-entrancy poses a significant problem for the state-of-the-art
analysis tools.
Securify and Mythril do not consider \eInst{CREATE} as an external call and
thus do not flag subsequent state updates.
Similarly, Oyente, Manticore, and ECFChecker consider only \eInst{CALL}
instructions when checking for re-entrancy vulnerabilities. Hence, they all
fail to detect create-based re-entrancy attacks.
Similar to delegated re-entrancy, the create-based re-entrancy vulnerability
emerges only when two contracts are combined. Thus, the contracts must be also
analyzed in combination, which is challenging as the contract code might change
after the analysis.

%% file: figures/cross-func-listings.tex
\begin{lstlisting}[language=solidity]
mapping (address => uint) tokenBalance;
mapping (address => uint) etherBalance;
 
function withdrawAll() public {
  uint etherAmount = etherBalance[msg.sender];
  uint tokenAmount = tokenBalance[msg.sender];
  if (etherAmount > 0 && tokenAmount > 0) {
    uint e = etherAmount + (tokenAmount * currentRate);
    etherBalance[msg.sender] = 0;
    // cannot re-enter withdrawAll()
    // However, can re-enter transfer()
    msg.sender.call.value(e)();   
    // state update causing inconsistent state
    tokenBalance[msg.sender] = 0;
  }
}
function transfer(address to, uint amount) public {
  // uses inconsistent tokenBalance (>0) when re-entered
  if (tokenBalance[msg.sender] >= amount) {
    tokenBalance[to] += amount;
    tokenBalance[msg.sender] -= amount;
  }
}
\end{lstlisting}

%% file: figures/delegated-listing.tex
\begin{lstlisting}[language=solidity]
library Lib {  // Library contract
  function send(address to, uint256 amount) public {
    to.call.value(amount)(); // CALL
  }
  // ...
}
contract Victim {
  mapping (address => uint) public credit;
  Lib lib; // address of library contract
  // ...
  function withdraw(uint amount) public {
    if (credit[msg.sender] >= amount) {
      // DELEGATECALL into Library
      address(lib).delegatecall(
        abi.encodeWithSignature("send(address,uint256)",
                                to, amount));
      // state update after DELEGATECALL
      credit[msg.sender] -= amount;
    }
  }
// ...
\end{lstlisting}

%% file: sections/overview.tex

In this section, we devise a novel way to detect re-entrancy attacks based on run-time
monitoring at the level of \evm\ bytecode instructions. Our approach,
called \toolname\ (Secure Ethereum), is based on extending an existing
Ethereum client, which we extend to perform run-time monitoring of contract
execution.

\begin{figure}[t]
  \centering
  \includegraphics[width=0.8\linewidth]{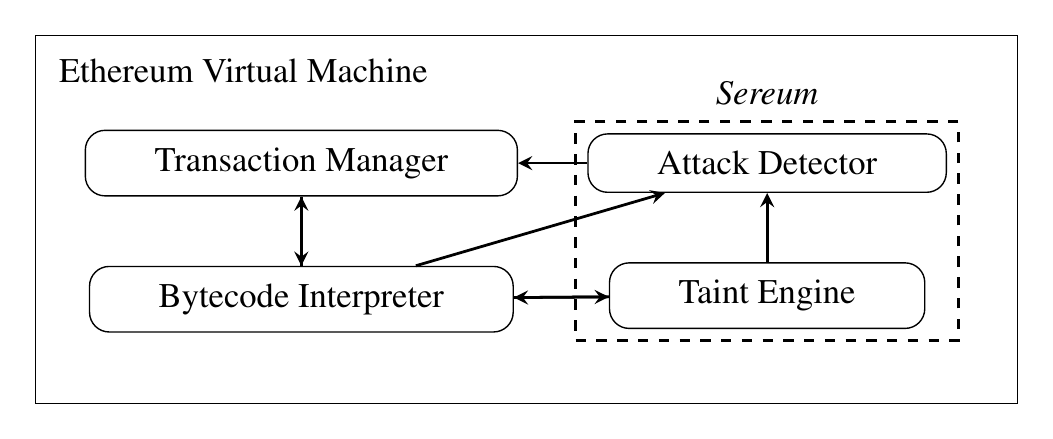}
  \caption{Architecture of enhanced \evm\ with run-time monitoring.}
  \label{fig:architecture}
\end{figure}

\noindent
\textbf{Architecture.}
Figure~\ref{fig:architecture} shows an overview of the \toolname\ architecture.
For a standard Ethereum client, the \evm\ features
a bytecode interpreter, which is responsible for executing the code of the
smart contracts,
and the transaction manager
that executes, verifies and commits new and old
transactions.
\toolname\ extends the \evm\ by introducing two
new components: (1)~a \emph{taint engine}, and (2)~an \emph{attack detector}.
The taint engine performs dynamic taint-tracking; dynamic taint tracking assigns labels to data at pre-defined sources
and then observes how the labeled data affects the execution of the
program~\cite{sok-taint-analysis}. To the best of our knowledge \toolname\ is
the first dynamic taint-tracking solution for smart contracts.
The attack detector utilizes the taint engine to recognize suspicious states of program
execution indicating that a re-entrancy attack is happening in the current transaction.
It interfaces with the transaction manager of the \evm\ to
abort transactions as soon as an attack is detected.

\noindent
\textbf{Detecting Inconsistent State.}
To effectively reason about a malicious re-entrance into a contract, we need to
detect whether a contract acts on inconsistent internal state (cf.
Figure~\ref{fig:simpledaowithdraw}). Note that any persistent internal state
is stored in the storage memory region of the \evm\ (cf.
Section~\ref{sec:background}). Variables which are
shared between different invocations of a contract are always stored in
the \emph{storage} region. As such, only the storage region is relevant for
re-entrancy detection.
Thus, \toolname\ applies taint tracking to storage variables as these are the
only internal state variables capable of affecting a contract's control flow in
a subsequent (re-entered) invocation of the contract. That said, only if a
control-flow decision is dependent on storage variables, an attacker can
manipulate the outcome of a conditional branch decision by re-entering the
contract and thereby manipulate the behavior of the contract.
Hence, re-entrancy attacks only apply to contracts that execute conditional
branches dependent on persistent internal state, i.e., the storage region.

The main idea behind \toolname\ is to detect state updates, i.e., altering of
storage variables, after a contract (denoted as \victim\ contract) calls into
another contract (denoted as \attacker\ contract).  Notice that not all state
updates resemble malicious behavior, but only those where \victim\ \emph{is
re-entered and acts upon the updated state}.
Typically, the goal of re-entrancy attacks is to bypass validity checks in the
business logic of the \victim\ contract. As such, \toolname\ focuses only on
conditional jumps and the data that influences the conditional jumps.
Notice that it is also possible for a contract to transfer Ether without
performing any validity check.  Obviously, deploying such a contract would be
highly dangerous and inefficient due to unnecessary consumption of gas. Hence,
we do not explicitly capture such cases in \toolname.
However, \toolname\ can be easily extended to cover this kind of re-entrancy
attack by issuing write-locks not only for behavior-changing variables, but
also for variables that are passed to other contracts during external calls
(such as Ether amount or call input).

\begin{figure}[t]
  \centering
  \includegraphics[width=0.9\linewidth]{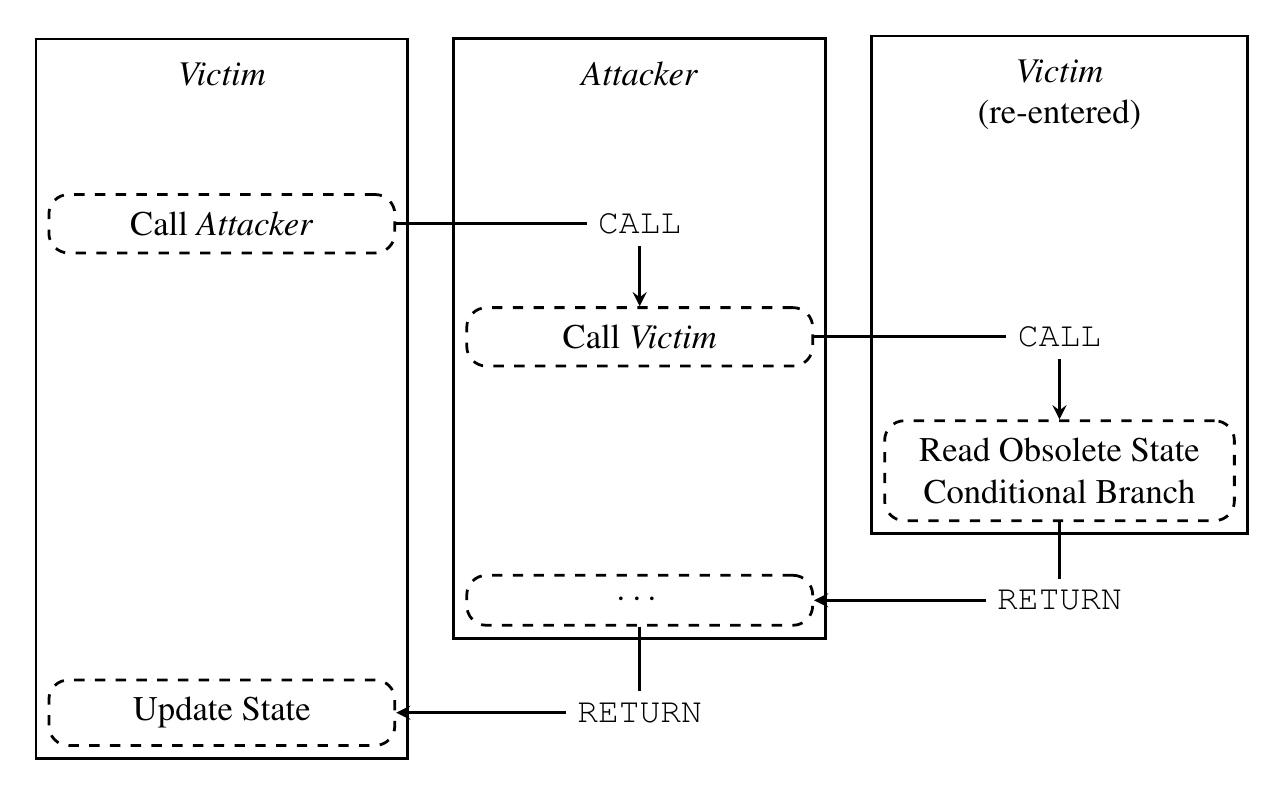}
  \caption{Re-entrancy attack exploits inconsistent state among different
  invocations of a contract.}
  \label{fig:inconsistentstate}
\end{figure}

Consider the example shown in Figure~\ref{fig:inconsistentstate},
\victim\ calls into the \attacker\ contract. The \attacker\ then forces a
re-entrancy into the \victim\ contract by calling into the \victim\ again. The
second re-entered invocation of \victim\ reads from a storage variable and
takes a control-flow decision based on that variable. After the \attacker\ contract
eventually returns again to \victim, the \victim\ contract will update the
state. However, at this point, it is clear that the re-entered \victim\ used a
wrong value read from inconsistent internal state for its conditional branch
decision.

The key observation is that inconsistent state can only arise if (1)~a
contract executes an external call to another contract, (2)~the storage
variable causing inconsistency is used during the external call for a
control-flow decision and (3)~the variable is updated after the external call returns.
Next, we describe in more details how the taint engine and the attack detector
detect inconsistent state at the \evm\ level.

%



\noindent
\textbf{Taint Engine and Attack Detector.}
To detect state updates, which cause inconsistency, we need to know which
storage variables were used for control-flow decisions.
On the \evm\ bytecode level a smart contract implements any control-flow decision as a conditional
jump instruction. Consequently, we leverage our taint engine to detect any
data-flow from a storage load to the condition processed by a conditional jump
instruction. This ensures that we only monitor those conditional
jumps which are influenced by a storage variable.
For every execution of a smart contract in a transaction, \toolname\ records
the set of storage variables, which were used for control-flow decisions.
Using this information, \toolname\ introduces a set of locks which prohibit
further updates for those storage variables.
If a previous invocation of the contract attempts to update one
of these variables, \toolname\ reports a re-entrancy problem and aborts the
transaction to avoid exploitation of the re-entrancy vulnerability.

In the simplest case, the attacker directly re-enters the victim contract.
However, the attacker might try to obfuscate the re-entrant call by first calling
an arbitrary long chain of nested calls to different attacker-controlled
contracts. Furthermore, during the external call, the attacker can re-enter the
victim contract several times, possibly in different functions (as shown in
the cross-function re-entrancy attack described in Section~\ref{sec:problem}).
This has to be taken into account when computing the set of locked storage variables.
To tackle these attacks,
\toolname\ builds a dynamic call tree during the execution of a transaction.
Every node in the dynamic call tree, represents a call to a contract and the
depth of the node in the tree is equal to the depth of the contract invocation
in the call stack of the \evm. We store those storage variables which influence
control-flow decisions as set $D_i$ for every node $i$ in the dynamic call tree.
The set of storage
variables $L_i$ that are locked at node $i$ is the union of $D_j$ for any node
$j$ of the same contract as $i$ that belongs to the sub-tree spanning from
node~$i$.

\begin{figure}[t]
  \centering
  \includegraphics[width=0.8\linewidth]{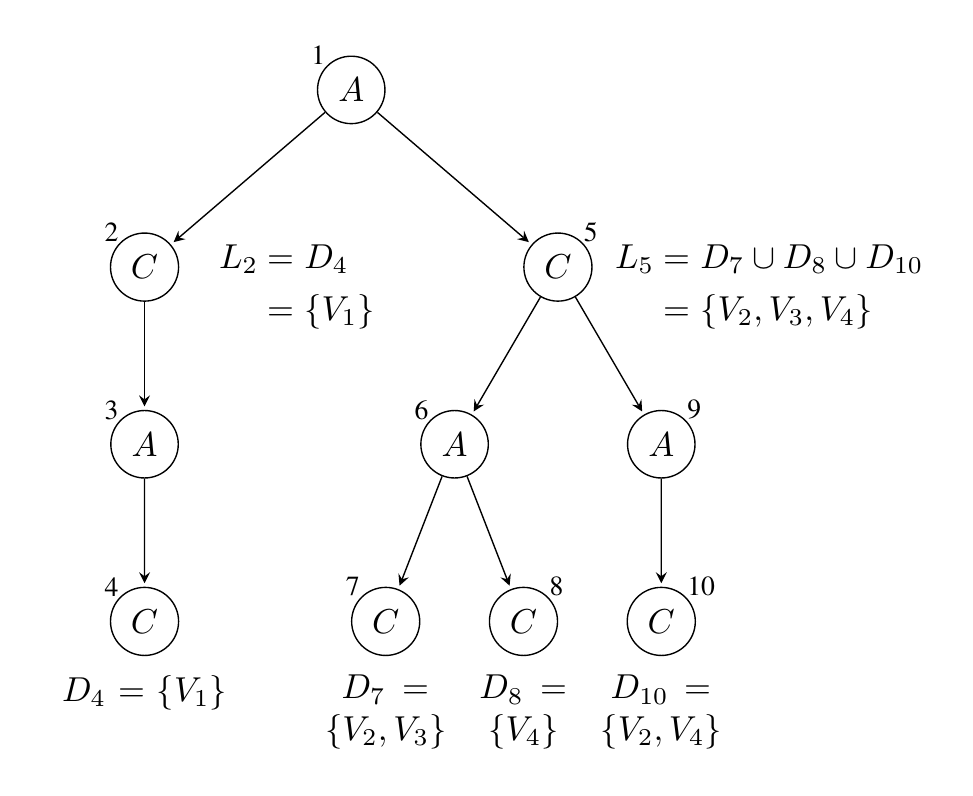}
  \caption{Dynamic call tree of a Ethereum transaction. Contract $A$ is
    re-entered several times. $V_k$ are storage variables. $D_i$ is the set of
    storage variables, which influence control-flow decisions in node $i$.
    $L_i$ is the set of storage variables, which are locked at node $i$ and cannot be
    updated anymore.}
  \label{fig:dyncallgraph}
\end{figure}

\noindent
\textbf{Example of Dynamic Call Tree.}
Figure~\ref{fig:dyncallgraph} depicts an example for
\toolname's generation of a dynamic call tree for a given Ethereum transaction. A possibly malicious
contract $A$ re-enters a vulnerable contract $C$ multiple times at different entry points
(functions). First, as shown on the left sub-tree, contract $A$ calls $C$,
$C$ calls $A$, and $A$ finally re-enters $C$. This sub-tree would be equivalent
to a classical re-entrancy attack, as shown previously in
Figure~\ref{fig:simpledaowithdraw}.
The variables locked during the first execution of contract $C$ (node marked
with $2$) are impacted only by the lower nodes in the call tree. The second execution
of contract $C$ (in node $4$) uses the storage variable $V_1$ for deciding a
conditional control-flow. Hence, this variable must not be modified after
the call in the execution of node $2$.

In contrast, the right side of the call
tree contains a more diverse set of nodes. For instance, the right part of the
call-tree could be part of a cross-function re-entrancy attack. We can observe
that different functions were called in the various re-entrant invocations of $C$, because
the variables used for conditional branches are different. Note that none of
the sets $D_5$, $D_7$, $D_8$, and $D_{10}$ are equal.
Contract $C$ performs two calls into
$A$ in node $5$. These calls re-enter $C$ in nodes $7$, $8$, and $10$. For
the execution of C from node $5$, we lock all variables from the sub-calltree below node
$5$.
Note that although variable $V_1$ is locked in node $2$, it is not in the set of
locked variables $L_5$. This means that no further calls starting from node $5$
have used the variable $V_1$ for a control-flow decision; thus $V_1$ can be
safely updated in node $5$, which will not change the behavior in any of the
nodes $7$, $8$ and $10$ unexpectedly.

A naive implementation of \toolname\ could just lock all variables which were used for
control-flow decisions. However, as we can see from
Figure~\ref{fig:dyncallgraph}, this would result in unnecessary locking of
variables when complex transactions are executed. This would also result in a
high number of false positives. For example, contract $C$ can safely update the
state variables $V_2$, $V_3$, and $V_4$ in node $2$, because they were not used
for conditional branches during the execution of node $4$. Similarly, node $5$
can safely update $V_1$ even though it was used for a control-flow decision in
a re-entrant call at node $4$.

The dynamic call tree allows \toolname\ to tackle the challenging new re-entrancy attacks we developed in Section~\ref{sec:problem}.
Recall that detecting \emph{cross-function re-entrancy} is challenging for static
analysis tools due to potential state explosion. Since \toolname\ performs dynamic
analysis, it does not suffer from such kind of weakness; it only analyzes those cross-function re-entrant calls that
actually occur at run-time.
Similarly, \emph{delegated re-entrancy} attacks are detected as \toolname\ --in contrast to existing tools-- does not inspect contracts in isolation, but analyzes and monitors exactly the library code
which is invoked when a transaction executes. 
That is, as an extension to the Ethereum client, \toolname\ can easily access
the entire blockchain state and hence retrieve the code of
every invoked library contract. Our taint engine simply propagates the
taints through the library. This also naturally covers any future updates of the library code.
Next, we describe the implementation details of \toolname.

%% file: sections/implementation.tex

We implemented \toolname\ based on the popular
\emph{go-ethereum}\footnote{\url{https://github.com/ethereum/go-ethereum}, based
on git commit 6a2d2869f6cb369379eb1c03ed7e55c089e83dd6/v1.8.3-unstable} project, whose client
for the Ethereum network is called \emph{geth}. In our implementation, we
extended the existing \evm\ implementation to include the taint engine and the
re-entrancy attack detector.
%
We faced one particular challenge in our implementation:
variables stored in the \emph{storage} memory region are represented
on the \evm\ bytecode level as load and store instructions to certain
addresses, i.e., any type information is lost during compilation. Hence,
only storage addresses are visible on the \evm\
level. Most storage variables, such as
integers, are associated with one address in the storage area. However, other
types, such as mapping of arrays, use multiple (not necessarily) adjacent
storage addresses. As such, \toolname\ tracks data-flows and sets the write-locks on
the granularity of storage addresses.

In the remainder of this section, we describe how \toolname\ tracks taints from
storage load instructions to conditional branches to detect storage addresses that
reference values that affect the contract's control-flow.
Furthermore, we show how \toolname\ performs attack detection by building the
dynamic call tree and propagating the set of write-locked storage addresses.

%% file: sections/taint.tex

%
Taint tracking is a popular technique for analyzing data-flows in
programs~\cite{sok-taint-analysis}. First, a \emph{taint} is assigned to a
value at a pre-defined program point, referred to as the so-called taint
source. The taint is propagated throughout the execution of the program along
with the value it was assigned to. Taint sinks are pre-defined points in the
program, e.g., certain instructions or function calls. If a tainted value
reaches a taint sink, the \toolname\ taint engine will issue a report, and
invoke the attack detection module.
Taint analysis can be used for both static and dynamic data-flow analysis.
Given that we aim to achieve run-time monitoring of smart contract, we leverage
dynamic taint tracking in \toolname.

To do so, we modified the bytecode interpreter of \emph{geth} ensuring that it
is completely transparent to the executed smart contract.
Our modified bytecode
interpreter maintains shadow memory to store taints separated from the actual
data values, which is a common approach for dynamic taint analysis.
%
\toolname\ allocates shadow memory for the different types of mutable memory in
Ethereum smart contracts (see Section~\ref{sec:evm}). The stack region can be
addressed at the granularity of 32-byte words. Thus, every stack slot is
associated with one or multiple taints. The storage address space is also
accessed at 32-byte word granularity, i.e., the storage can be considered as a
large array of 32-byte words, where the storage address is the index into that
array. As a result, we treat the storage region similar to the stack and
associate one or multiple taints for every 32-byte word. However, unlike
 the stack and storage address space, the memory region can be accessed at byte
granularity. Hence, we associate every byte in the memory address space with
one or multiple taints.
To reduce the memory overhead incurred by the shadow memory for the memory
region, we store taints for ranges of the memory region. For example, if the
same taint is assigned to memory addresses 0 to 32, we only store one taint for
the whole range. When only the byte at address 16 is assigned a new taint, we
split the range and assign the new taint only to the modified byte.

We propagate taints through the computations of a smart contract. As a general
taint propagation rule for all instructions, we take the taints of the input
parameters and assign them to all output parameters.
Since the \evm\ is a stack machine, all instructions either use the stack to
pass parameters or have constant parameters hard-coded in the code of the
contract. Hence, for all of the computational instructions, such as arithmetic
and logic instructions, the taint engine will pop the taints associated with
the instruction's input parameters from the shadow stack and the output of the
instruction is then tainted with the union of all input taints. Constant
parameters are always considered untainted. This ensures that we capture
data-flows within the computations of the contract.
One exception is the \eInst{SWAP} instruction family, which swaps two items on
the stack. The taint engine will also perform an equivalent swap on the shadow
stack without changing taint assignments.
Whenever a value is copied from one of the memory areas to another area, we also
copy the taint between the different shadow areas. For instance, when a value is
copied from the stack to the memory area, i.e., the contract executes a
\eInst{MSTORE} instruction, the taint engine will pop one taint from the shadow
stack and store it to the shadow memory region.
The \evm\ architecture is completely deterministic; smart contracts in the
\evm\ can only access the blockchain state using dedicated instructions. That is, no other
form of input or output is possible. This allows us to completely model the
data-flows of the system by tracking data-flows at the \evm\ instruction level.

For re-entrancy detection, as described in
Section~\ref{sec:overview}, we only need one type of taint,
which we call
\emph{DependsOnStorage}.
The taint source for this taint is the \eInst{SLOAD}
instruction. Upon encountering this instruction, the taint engine creates a
taint, which consists of the taint type and the address passed as operand to
the \eInst{SLOAD} instruction.
The conditional \eInst{JUMPI} instruction is used as a taint
sink. Whenever such a conditional jump is executed, the taint engine checks
whether the condition value is tainted with a \emph{DependsOnStorage} taint.
If this is the case, the taint engine will extract the storage address from the
taint and add it to the set of variables that influenced control-flow
decisions. Our implementation supports an arbitrary number of
different \emph{DependsOnStorage} taints. This allows \toolname\ to support
complex code constructs, e.g., control-flow decisions which depend on multiple different storage variables.

\begin{figure*}[ht]
  \centering
  \includegraphics[width=\linewidth]{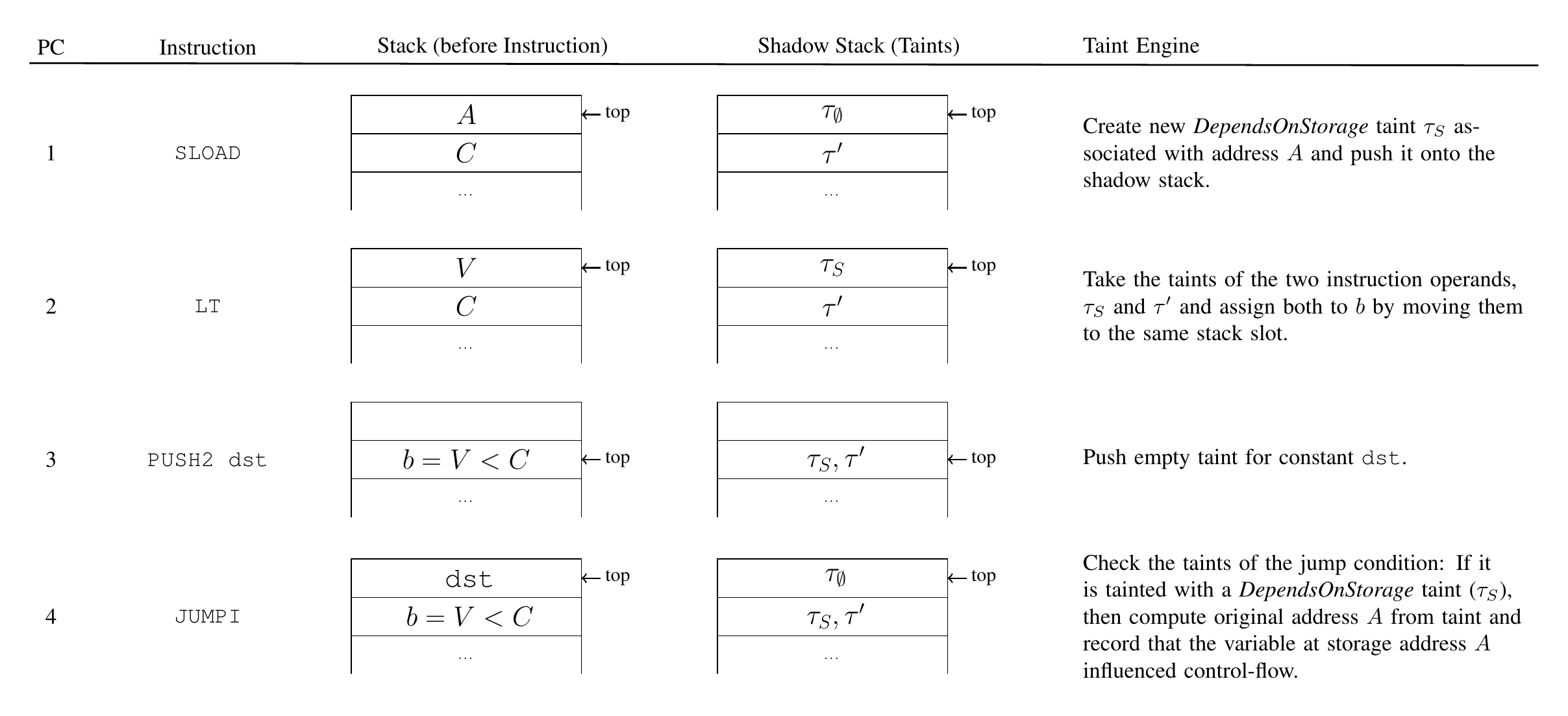}
  \caption{The taint engine propagates the taints $\tau$ through the executed
    instructions and stores them on a shadow stack. The condition for the
    conditional jump $b$ depends on the values $C$ and the value $V$, which was
    loaded from storage address $A$. $SP$ is the current stack pointer,
    pointing to the top of the data stack.}
  \label{fig:taintprop}
\end{figure*}

\noindent
\textbf{Example for Taint Assignment and Propagation.}
Figure~\ref{fig:taintproplisting} shows a snippet of Ethereum bytecode
instructions. In this snippet of instructions, there exists a data-flow from
the \eInst{SLOAD} instruction in line 1 to the conditional jump instruction in
line 4. The \eInst{SLOAD} instruction will load a value from the storage memory
region. The first and only parameter to \eInst{SLOAD} is the address in the
storage area. The \eInst{JUMPI} instruction takes two parameters: the jump
destination and the condition whether the jump is to be performed. Recall that
all instruction operands except for the \eInst{PUSH} instruction are passed via the stack.
Figure~\ref{fig:taintprop} shows the state of the normal data stack and the
corresponding shadow stack, when the snippet in Figure~\ref{fig:taintproplisting} is
executed. $SP$ denotes the stack pointer before the instruction is executed.
The \eInst{SLOAD} instruction will pop an address $A$ from the stack, load the
value $V$ (referenced by $A$) from storage, and then push it onto the stack. Since, the
\eInst{SLOAD} instruction is defined as taint source, the taint engine will create a new
\emph{DependsOnStorage} taint, which we denote as $\tau_s$. This taint is
assigned to the value $V$ by pushing it onto the shadow stack. Note that in
this case $V$ was not previously assigned a taint.
The instruction
\eInst{LT} (\emph{less-than}) compares the value loaded from storage with the
value $C$ that was previously pushed on the stack. This comparison decides
whether the conditional jump should be taken. Since the \eInst{LT} instruction
takes two parameters from the stack ($V$ and $C$), the taint engine also pops
two taints from the shadow stack ($\tau_s$ and $\tau'$). The result of the
comparison is then tainted with both taints ($\tau_s$ and $\tau'$), so the
taint engine pushes a merged taint ($\tau_s , \tau'$) to the shadow stack.
The \eInst{PUSH2} instruction then pushes a 2-byte constant to the stack, which
is assigned an empty taint $\tau_\emptyset$. Finally, the \eInst{JUMPI}
instruction takes a code pointer (\texttt{dst}) and a boolean condition as
parameters from the stack. Since \eInst{JUMPI} is a taint sink, the taint
engine will check the taints associated with the boolean condition. If this
value is tainted with the $\tau_S$ taint, it will compute the original storage
address $A$ based on the taint. At this point, we know that the value at storage address
$A$ influenced the control-flow decision. Hence, we add it to the set of control-flow influencing
storage addresses, which is passed to the attack detection component later on.

\begin{figure}[ht]
  \centering
  \input{figures/taintprop-listing}
  \caption{Ethereum assembly snippet implementing a solidity
    \emph{if-statement} with a conditional branch.
    The \eInst{SLOAD} Instruction in line 1 indirectly influences the
    control-flow decision in the \eInst{JUMPI} instruction in line 4 as it is
    used as a parameter in the  \eInst{LT} instruction. \eInst{LT}
    performs a \emph{less-than} comparison between the first and
  second operand on the stack.}
  \label{fig:taintproplisting}
\end{figure}

Using the taint engine, \toolname\ records the set of storage addresses that reference
values which influence control-flow decisions. This set of addresses
is then forwarded to the attack detection component once the contract finishes
executing.

%% file: figures/taintprop-listing.tex
\begin{lstlisting}[language=,escapechar=!]
SLOAD
LT
PUSH2 dst
JUMPI
\end{lstlisting}

%% file: sections/reentrancy.tex

%
To detect re-entrancy attacks, we lock the write-access to storage
addresses that influence control-flow decisions. During execution of a
contract, the taint engine detects and records storage addresses, which are
loaded and then influence the outcome of a control-flow decision.
As described in Section~\ref{sec:overview}, \toolname\ uses a dynamic call-tree
to compute the set of variables that are locked for writing.
\toolname\ builds the dynamic call-tree during execution of a transaction.
This tree contains a node for every invocation of a contract during the
transaction.  The dynamic call-tree records how the call stack of the
transaction evolves over time. The ordering of the child nodes in the dynamic
call-tree corresponds to the order of execution during the transaction. The
depth of the node in the tree corresponds to the depth in the call stack, i.e., the time when
a contract was invoked. The dynamic call-tree is updated whenever a contract
issues or returns from an external call.
When the called contract completes
execution, the set of control-flow influencing variables is retrieved from the
taint engine and stored in the node of the call-tree.

\toolname\ locks only the set of variables, which were used for control-flow
decisions during an external call. To compute this set, \toolname\ traverses
the dynamic call-tree starting from the node corresponding to the current
execution. During traversal, \toolname\ searches for nodes, which were part of
executions of the same contract. When \toolname\ finds such a node, it retrieves
the set of control-flow influencing variables previously recorded by the taint
engine.
\toolname\ updates the set of locked variables after every external call.
Whenever a contract attempts to write to the storage area, i.e., executes the
\eInst{SSTORE} instruction, \toolname\ intercepts the write and first checks
whether the address is locked. If the variable is locked, \toolname\ reports
a re-entrancy attack and then aborts execution of the transaction.
This results in the \evm\ unwinding all state changes and Ether transfers.

%% file: sections/evaluation.tex

In this section, we evaluate the effectiveness and performance of \toolname\
based on existing Ethereum contracts deployed on the Ethereum \emph{mainnet}.
Since our run-time analysis is transparently enabled for each execution of a
contract, we re-execute the transactions that are saved on the Ethereum
blockchain.  We compare our findings with state-of-the-art academic analysis
tools such as Oyente~\cite{luu2016making, oyente-tool} and
Securify~\cite{securifych}. Note that we do not compare with
Zeus~\cite{zeus-ndss2018} and SmartCheck~\cite{smartcheck} since these require
access to the source code of contracts which is rarely available for existing
contracts.
The latest version of Securify, which is only available through a web
interface, does not support submitting bytecode contracts anymore. Therefore,
we were not able to test all contracts with Securify.
Furthermore, we do not compare with Mythril~\cite{mythril} and
Manticore~\cite{manticore} as they follow the detection approach of Oyente
(symbolic execution).
%
%
We also conduct experiments based on the three new re-entrancy attack patterns
we introduced in Section~\ref{sec:problem}---effectively demonstrating that
only \toolname\ is able to detect them all.

\subsection{Run-time Detection of Re-Entrancy Attacks}

We first connect our \toolname\ client with the public Ethereum network to retrieve all
the existing blocks while keeping as many intermediate states in the cache as
possible. Transaction re-execution requires the state of the context
block. States are saved as nodes in the so-called state Patricia tree of the
Ethereum blockchain. We run the \emph{geth} (Go Ethereum)
client with the options sync mode \emph{full}, garbage collection
mode \emph{archive}, and assign as much memory as possible for the cache.
During the block synchronization process, the taint tracking option of
\toolname\ is disabled to ensure that the client preserves the original state at each
block height.

We then replay the execution of each transaction in the blockchain.
To reduce the execution time, we limit our testset until block
number 4,500,000.
Note that we skip those blocks which were target of
denial-of-service attacks as they incur high execution times of transactions~\cite{ethdos}. We
replay the transactions using the \funcname{debug} module of the
\emph{geth} RPC API. This ensures that our replay of transactions does not affect the public saved blockchain data.
We also retrieve an instruction-level \emph{trace} of the executed instructions and the corresponding
storage values during the transaction execution. This allows us to step through
the contract's execution at the granularity of instructions.

We enable the taint tracking option in \toolname\ during the transaction
replay to evaluate whether a transaction triggers a re-entrancy attack
pattern; in this case, an exception will be thrown, the execution of the
transaction gets invalidated, and an error is reported via the API.
\toolname\ will then return the instruction trace up to the point where the
re-entrancy attack is detected.

All in all, we re-executed 77,987,922 transactions involved in these 4.5 million
blocks, and \toolname\ has flagged
49,080 (0.063\%) of them as re-entrancy violation. Originally, we identified 52 involved contracts that
count up to only 0.055\% of the total number of 93,942~\footnote{We
count the number of contracts created by transactions sent to the contract
creation address `0x0'. We do not count those contracts created by other
contracts, which will result in a higher number.}
created contracts in our testset.
However, while manually analyzing these contracts, we discovered that many contracts are created
by the same account and share the same contract code; they are only instantiated
with different parameters. As such, we consider these contracts as being \emph{identical}. More specifically, we found
three groups of identical contracts involving 21, 4, and 3 contracts, respectively.
Similarly, we identified that a number of contracts execute the same sequence of
instructions that only differ in the storage addresses.
We consider these contracts as \emph{alike} contracts. In total, we found two groups of alike
contracts of size 10 and 3, respectively.
As a result, \toolname\ detected 16 identical or alike contracts that are
invoked by transactions matching the re-entrancy attack pattern.

\begin{figure}[t]
  \centering
  \includegraphics[width=\linewidth]{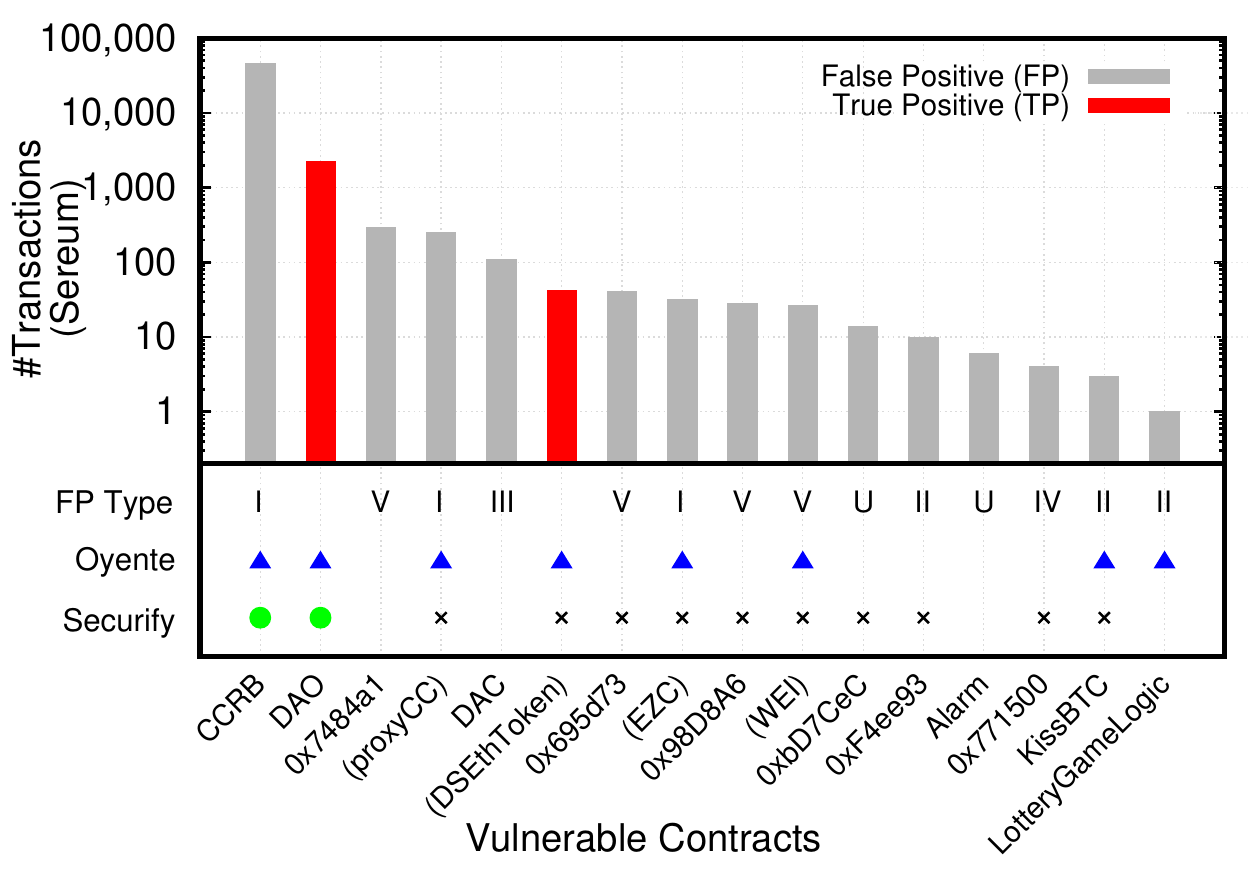}
  \caption{The top plot shows the number of detected transactions triggering the
    re-entrancy vulnerability in the flagged contracts. Each contract is
    categorized by its false positive type described in
    Section~\ref{sec:fpanalysis}.
    Type I corresponds to ``lack of field-sensitivity'',
    Type II ``storage deallocation'',
    Type III ``constructor callbacks'',
    Type IV ``tight contract coupling'',
    Type V ``manual re-entrancy locking'',
    and U for Unknown.
    The contract name is shown for those where source code available. Contracts in parenthesis are
    known token contracts at \href{etherscan.io}{http://etherscan.io} although source code
    is not available. The bottom plot shows how the tools
    Oyente~\cite{oyente-tool} and Securify~\cite{securifych} handle this subset of contracts. Since
    the last public version of Securify requires source code,
    we add a cross for those (bytecode) contracts we were not able to evaluate.}
  \label{fig:tx-dist}
\end{figure}

For 6 out of these 16
contracts, the source code is available on
\href{etherscan.io}{http://etherscan.io}, thus allowing us to perform detailed investigation why they have been flagged.
In what follows, we manually check whether a violating transaction resembles a real
re-entrancy attack, and whether the concerned contract suffers from re-entrancy
vulnerability that could potentially be exploited.

For contracts with Solidity source code, we perform source code review and
check the contract logic provided the transaction input to manually identify
re-entrancy attacks.
We use the transaction trace as a reference to follow the control flow
and observe the intra-contracts calls.
For contracts with no source code, we cannot fully recover the contracts semantics for
detected inconsistent state updates. In this case, we use
the transaction trace and the \emph{ethersplay}~\cite{ethersplay-tool}
disassembler tool
to partially reverse-engineer the contracts.

Based on our investigation, we can confirm that two contracts were
actually exploited by means of a re-entrancy attack. One of them is the known
\emph{DAO}~\cite{dao-contract} attack attributing to 2,294
attack transactions.\footnote{Note that we consider
\emph{TheDarkDAO}~\cite{darkdao-contract} and \emph{DAO}~\cite{dao-contract} contract as being identical.}
The second case involves a quite unknown re-entrancy attack. It occurred at contract address
\verb+0xd654bDD32FC99471455e86C2E7f7D7b6437e9179+ and attributed to 43 attack
transactions. After reviewing blog posts and GitHub repositories related to this contract~\cite{makerdao-reddit,makerdao-hack-recovery},
we discovered that this contract is known as \emph{DSEthToken} and is
part of the \emph{maker-otc} project.
This series of attack transactions were initiated by
the contract developers after they discovered a re-entrancy vulnerability.
Since the related funds were drained by (benign) developers,
the Ethereum community payed less attention to this incident.
In total, \toolname\ incurs a false positive rate as low as 0.06\% across all
the re-run transactions.
Figure~\ref{fig:tx-dist} shows the number of transactions that match the
re-entrancy attack pattern flagged by \toolname. Some of the results reflect
false positives which will be discussed in detail in Section~\ref{sec:fpanalysis}.


We also observe that Oyente flagged 8 of these contracts as vulnerable to
re-entrancy attacks. Some contracts were not detected by Oyente since Oyente
does not consider any of the advanced re-entrancy attacks we discussed in
Section~\ref{sec:problem}.  During our analysis we noticed that in some cases
Oyente warned about re-entrancy problems, which are only exploitable with a
cross-function re-entrancy attack. However, we believe this is due to Oyente
incorrectly detecting a same-function re-entrancy vulnerability.
%
%
Apart from the 6 false positives in our test set, the analysis performed by
previous work~\cite{securifych,maffei-staticanalysis} demonstrated that
re-entrancy detection in Oyente suffers from false positive issues.

With respect to Securify, the latest version of Securify requires the source code of a contract
thereby impeding us from evaluating all contracts. We therefore have only examined the
contracts whose source code is available.
Securify defines a very conservative
violation pattern for re-entrancy detection that forbids any state update after
an external call. As such, 5.8\% out of 24,594 tested contracts in the authors'
experiment (around 1,426 contracts) are flagged as vulnerable to re-entrancy, which consequently
results in a very high false positive rate.

Lastly, we evaluated our new re-entrancy attack patterns
(Section~\ref{sec:problem}).  For each contract, we crafted one attack
transaction for \toolname\ to perform the check: \toolname\ successfully
detects all attack transactions to the three vulnerable contracts.
Table~\ref{tab:toolseval} shows an overview of various tools tested against the
vulnerable contracts for the new re-entrancy attacks patterns.
As discussed earlier, neither Oyente, Securify nor Manticore were able to detect delegated and
create-based re-entrancy vulnerabilities. While Oyente does not detect the
cross-function re-entrancy attack, Securify is able to detect it due to its
conservative policy. Similarly, Mythril detects cross-function and create-based
re-entrancy, because it utilizes a similar policy to Securify, which is
extremely conservative and therefore also results in a high number of false
positives.
ECFChecker detects the cross-function re-entrancy attack. However, during our
evaluation, we crafted another contract, which is vulnerable to cross-function
re-entrancy, but was not detected by ECFChecker.
Recall that the delegated re-entrancy attack cannot be detected by any existing static
off-line tool as it exploits a dynamic library which is either not available at
analysis time or might be updated in the future. However, a dynamic tool, such
as ECFChecker, can detect the delegated re-entrancy.
The create-based re-entrancy attack is not detected by any of the existing
analysis tools, as the instruction \eInst{CREATE} is currently not considered
as an external call by none of the existing analysis tools.

\begin{table}[b]
  \input{figures/tools_eval}
\end{table}

In general, we argue that \toolname\ offers the advantage of detecting actual re-entrancy
attacks and not possible vulnerabilities.
As such, we can evaluate on a reduced set of only 16 contracts rather than 185 (Oyente) or 1,426 (Securify) contracts.
In contrast to previous work \cite{luu2016making,securifych}, this makes it
feasible for us to exactly determine whether an alarm is a true or false positive.
Moreover, some of the contracts are not flagged by Oyente and Securify as these do not cover the full
space of re-entrancy attacks. As such, they naturally do not raise false positives for contracts that violate
re-entrancy patterns that are closely related to the delegated and create-based re-entrancy (i.e., Type~III and~IV).

\subsection{False Positive Analysis}
\label{sec:fpanalysis}
\input{sections/falsepositiveanalysis}

\subsection{Performance and Memory Overhead}

\input{sections/performance}

%% file: figures/tools_eval.tex

\newcommand{\myYes}{\CIRCLE}
\newcommand{\myNo}{\Circle}
\newcommand{\myPartial}{\LEFTcircle}

\centering
\caption{Comparison of re-entrancy detection tools subject to our
  testcases for the advanced re-entrancy attack patterns. Tools marked with
  \myYes\ support detecting this type of re-entrancy, while tools marked with
  \myNo\ do not support detecting this type of re-entrancy. Tools with an
  overly restrictive policy are marked with \myPartial.}
\label{tab:toolseval}
\begin{tabular}{llcccc}
  \toprule
  Tool & Version & Cross-Function & Delegated & Create-based \\
  \midrule
  Oyente     & 0.2.7       & \myNo & \myNo & \myNo \\
  Mythril    & 0.19.9      & \myPartial & \myNo & \myPartial \\
  Securify   & 2018-08-01  & \myPartial & \myNo & \myNo \\
  Manticore  & 0.2.2       & \myYes & \myNo & \myNo \\
  ECFChecker & geth1.8port & \myYes & \myYes & \myNo \\
  Sereum     & -           & \myYes & \myYes & \myYes \\
\bottomrule
\end{tabular}

%% file: sections/falsepositiveanalysis.tex

While investigating the 16 contracts which triggered the re-entrancy
detection of \toolname, we discovered code patterns in deployed contracts (see
Figure~\ref{fig:tx-dist}),
which are challenging to accurately handle for any off-line or run-time bytecode analysis tool.
These patterns are the root cause for
the rare false positive cases we encountered during our evaluation of \toolname.

However, since these code patterns are not only challenging for \toolname, but for
other existing analysis tools such as Oyente~\cite{oyente-tool},
Mythril~\cite{mythril}, Securify~\cite{securifych}, or any reverse-engineering tools operating at \evm\ bytecode level~\cite{erays,ethersplay-tool},
we believe that a detailed investigation of these cases is highly valuable
for future research in this area. Our investigation also \emph{reveals for the first time} why existing tools
suffer from false alarms when searching for re-entrancy vulnerabilities. In what follows, we reflect on the investigation of the false positives that we encountered.

\paragraph{\rom{1}. Lack of Field-Sensitivity on the \evm\ Level.}
Some false positives are caused by lack of
information on fields at bytecode level for data structures.
Solidity supports the keyword \texttt{struct} to
define a data structure that is composed of multiple types, e.g.,
Figure~\ref{fig:soliditystruct} shows a sample definition of a \texttt{struct}
$S$ of size 32 bytes.
Since the whole type can be stored within one single word in the \evm\ storage
area, accessing either of the fields $a$ or $b$ ends up accessing the same storage address.
In other words, on the \evm\ bytecode level, the taint-tracking engine of
\toolname\ cannot differentiate the access to fields $a$ and $b$.
This leads to a problem called \emph{over-tainting}, where taints spread to
unrelated values and in turn causes false positives.
%
Notice that this problem affects all analysis tools working on the \evm\ bytecode level.
Some static analysis tools~\cite{porosity} use heuristics to detect the high-level types in
Ethereum bytecode. The same approach could be used to infer the
types of different fields of a packed data structure. However, for a run-time
monitoring solution, heuristic approaches often incur unacceptable runtime
overhead without guarantee of successful identification.
To address this type of false positive, one would either
require the source code of the contract or additional type information on the
bytecode level.

\begin{figure}[t]
  \centering
  \input{figures/field-sensitivity-listing}
  \caption{Solidity \emph{struct}, where both \eVar{a} and \eVar{b} are at the
  same storage address. Therefore, any update to \eVar{a} or \eVar{b} includes
  loading and writing also the other.}
  \label{fig:soliditystruct}
\end{figure}

\paragraph{\rom{2}. Storage Deallocation.}
Recall that the \evm\ \emph{storage} area is basically a key-value store that maps 256-bit
words to 256-bit words.
The \evm\ architecture guarantees that the whole \emph{storage} area is
initialized with all-zero values and is always available upon request.
More specifically, no explicit memory allocation is required,
while memory deallocation simply resets the value to zero.
This poses a problem at the bytecode level: a memory deallocation is no
different from a state update to value $0$, though the semantics differ;
especially when applying the re-entrancy detection logic.
%
Consider the example of a map \emph{M} in Figure~\ref{fig:soliditydelete}.
%
When the contract deallocates the element indexed by $id$ from $M$ (\funcname{delete} from a
map), it basically has the same effect as setting the value of $M[id]$ to $0$ at
the bytecode level. Here, the Solidity compiler will emit nearly identical
bytecode for both cases.
%
We encountered a contract\footnote{Contract address:
0x6777c314b412f0196aca852632969f63e7971340} presenting this case which leads to a
false alarm. Similar to field-sensitivity issues, correctly handling such
cases requires the source code or an explicit \evm\ deallocation instruction.

\begin{figure}[t]
  \centering
  \begin{lstlisting}[language=solidity]
mapping (uint => uint) M; // a hash map
// delete entry from mapping
delete M[id];
// on the EVM level this is equivalent to
M[id] = 0;
\end{lstlisting}
  \caption{Solidity storage delete is equivalent to storing zero.}
  \label{fig:soliditydelete}
\end{figure}

\paragraph{\rom{3}. Constructor Callbacks.}
\toolname\ considers calls to the constructor of contracts to be the same as
calls to any other external contract. This allows \toolname\ to detect
create-based re-entrancy attacks (cf. Section~\ref{sec:attackcreatebased}).
However, detecting create-based re-entrancy comes at the cost of some false
positives.
During our evaluation\footnote{Contract address
0xFBe1C2a693746Ccfa2755bD408986da5281c689F}, we noticed that sub-contracts
created by other contracts, tend to call back into their parent contracts.
Usually, this is used to retrieve additional information from the parent
contract: the parent creates the sub-contract, the sub-contract re-enters
the parent contract to retrieve the value of a storage variable, and
that same variable is then updated later by the parent.
Consider the example in Figure~\ref{fig:constructorcallback}, where contract
$A$ creates a sub-contract $B$. While the constructor executes, $B$ re-enters the parent
contract $A$, which performs a control-flow decision on the \eVar{funds}
variable. This results in \toolname\ locking the variable \eVar{funds}.
Since no call to another potentially malicious external contract is involved
this example is not exploitable via re-entrancy.
However, \toolname\ detects that the \eVar{funds} variable is possibly
inconsistent due to the deferred state update. A malicious contract $B$ could have
re-entered $A$ and modified the \eVar{funds} variable in the meantime.

We argue that this constructor callback pattern should be avoided by contract
developers. All necessary information should be passed to the sub-contract's
constructor, such that no re-entrancy into the parent contract is needed. This
does not only avoid false positives in \toolname, but also decreases the gas costs.
External calls are one of the most expensive instructions in terms of gas
requirements, which must be payed for in Ether and as such should be avoided as
much as possible.

\begin{figure}[t]
  \centering
\begin{lstlisting}[language=solidity]
contract A {
  mapping (address => uint) funds;
  // ...
  function hasFunds(address a) public returns(bool) {
    // funds is used for control-flow decision
    if (funds[a] >= 1) { return true; }
    else { return false; }
  }
  function createB() {
    B b = new B(this, msg.sender);
    // ...
    // update state (locked due to call to hasFunds)
    funds[msg.sender] -= 1;
  }
}
contract B {
  constructor(A parent, address x) {
    // call back into parent
    if (parent.hasfunds(x)) { /* ... */ }
  }
}
\end{lstlisting}
  \caption{Constructor callback. The sub-contract $B$ calls back (re-enters) into the
           \eFunc{hasFunds} function of the parent contract $A$. This
           type of false positive is similar to the create-based re-entrancy
           attack pattern.}
  \label{fig:constructorcallback}
\end{figure}

\paragraph{\rom{4}. Tight Contract Coupling.}
%
During our evaluation, we noticed a few cases where multiple contracts are tightly
coupled with each other resulting in overly complex transactions, i.e., transactions
that cause the contracts to be re-entered multiple times into various functions. This
suggests that these contracts have a strong interdependency.
Since \toolname\ introduces locks for
variables that can be potentially exploited for re-entrancy and is not aware of the underlying
trust relations among contracts, it reports a false alarm
when a locked variable is updated.
We consider these cases as an example for bad contract development practice since
performing external calls is relatively expensive in terms of gas, and such
also Ether, and could be easily avoided in these contracts. That is, if trusted contracts have
internal state that depends on the state of other trusted contracts, we
suggest developers to keep the whole state in one contract and use
safe library calls instead.

\paragraph{\rom{5}. Manual Re-Entrancy Locking.}
To allow expected and safe re-entrancy, a smart contract can manually introduce
lock variables (i.e., a mutex) to guard the entry of the function. In
Figure~\ref{fig:manuallocks}), \eFunc{disableWithdraw} enables a lock at
\circletwo\ before making an external call at \circlethree. The lock is reset
after the call at \circlefour. This prevents any potential re-entrance at
\circleone.  Hence, even though the \emph{balance} is updated after the
external call, the contract is still safe from re-entrancy attacks.

\begin{figure}[t]
  \centering
  \input{figures/manual-locks}
  \caption{Manual locking to guard against re-entrancy.}
  \label{fig:manuallocks}
\end{figure}

However, the access pattern to these lock variables during contract re-entrance
matches an attack pattern, i.e., the internal state (the lock variable) that
affects the control flow in subsequent (re-entered) invocation of the contract,
is updated subsequently (at \circlefour). Operating at bytecode level, it is
challenging to distinguish the benign state updates of \emph{locks} from those
of critical variables such as \emph{balances}. Note that manual locking is an
error-prone approach as it could allow an attacker to re-enter other functions
of the same contract, unless the entry of every function is guarded by the
lock. In contrast, \toolname\ automatically introduces locks for all possibly
dangerous variables (detected via taint tracking) across all functions thereby
removing the burden from developers to manually determine all possible
vulnerable functions and critical variables.

%% file: figures/field-sensitivity-listing.tex

\begin{lstlisting}[language=solidity]
struct S {
  int128 a;  // 16 bytes
  int128 b;  // 16 bytes
}            // total: 32 bytes (one word in storage)
\end{lstlisting}

%% file: figures/manual-locks.tex
\begin{lstlisting}[language=solidity,escapechar=!]
mapping (address => uint) private balances;
mapping (address => bool) private disableWithdraw;
// ...
function withdraw() public {
!\circleds{1}! if (disableWithdraw[msg.sender] == true) { 
     // abort immediately and return error to caller
     revert(); 
   }
   uint amountToWithdraw = balances[msg.sender];
    
!\circleds{2}! disableWithdraw[msg.sender] = true;
!\circleds{3}! msg.sender.call.value(amountToWithdraw)();
!\circleds{4}! disableWithdraw[msg.sender] = false;
   // state update after call
   userBalances[msg.sender] = 0;
}
\end{lstlisting}

%% file: sections/performance.tex

Since there are no benchmarks, consisting of realistic contracts, available for
\evm\ implementations, we measured the performance overhead by timing the
execution of a subset of blocks from the Ethereum blockchain. We sampled blocks
from the blockchain, starting from $460000$, $450000$, $440000$, $4300000$ and
$4200000$, we use 10 consecutive blocks. We run those 50 blocks in batch
$10000$ times, while accounting only for the \evm's execution time. We perform
one run with plain \emph{geth}, on which \toolname\ is based, and one with
\toolname\ with attack detection enabled.
For the performance evaluation, we do not consider those transactions, which
\toolname\ flags as a re-entrancy attack. \toolname\ aborts those transactions
early, which can result in much shorter execution time, compared to the normal
execution.
%
%
We measured the performance overhead of \toolname, compared with plain
\emph{geth} when running 50 blocks in batch. Here, we average the runtime over 10,000
runs of the same 50 blocks. We benchmarked on a 8-core Intel(R) Xeon(R) CPU
E5-1630 v4 with 3.70GHz and 32 GB RAM. The mean runtime of \emph{geth} was
2277.0 ms ($\sigma =$ 146.7 ms). The mean runtime of \toolname\ was 2494.5 ms
($\sigma =$ 174.8 ms). As such, \toolname\ incurred a mean overhead of 217.6 ms
($\sigma =$ 100.9 ms) or $9.6\%$.
While measuring the timing of the executed transactions, we additionally measured the memory
usage of the whole Ethereum client. We used Linux cgroups to capture and
measure the memory usage of \toolname\ and all subprocesses. We sample the
memory usage every second while performing the runtime benchmarks. During our
benchmark, \toolname\ required on average 9767 MB of memory with active attack
detection, while the plain \emph{geth} required 9252 MB.
%

%
%
%
This shows that \toolname\ can effectively detect re-entrancy attacks with a
negligible overhead. In fact, the actual runtime overhead is not noticeable. The
average time until the next block is mined in $14.5$ seconds and contains 130
transactions on average (between Jan 1, 2018 until Aug 7, 2018). Given our
benchmark results, a rough estimate of \evm\ execution time per block is $0.05$
seconds, with \toolname\ adding $0.005$ seconds overhead. Compared to the
total block time the runtime overhead of \toolname\ is therefore not
noticeable during normal usage.
%


%% file: sections/relatedwork.tex

In this section, we overview related work in the area---beyond the state of the
art defenses and analysis tools that have been described in Section~\ref{sec:background}.

Vyper~\cite{vyper} is an experimental
language dedicated to maximize the difficulty of writing \emph{misleading} code
while ensuring human-readability to enable easy auditing of the contract. It
achieves better code clarity by considerably limiting high-level programming
features such as class inheritance, function overloading, infinite loops, and
recursive calls. This approach sacrifices the expressiveness of the language in
exchange for gas predictability.
\emph{Babbage}~\cite{babbage} has been recently proposed by the Ethereum community as
a visual programming language that consists of mechanical components aiming to help programmers to better understand the interactivity of components in a contract.
\emph{Bamboo}~\cite{bamboo} is another contract programming language focusing
on the state transition of contracts. A contract is described as a
\emph{state machine} whose state will change along with the contract signature.
\emph{Obsidian}~\cite{Coblenz2017-obsidian} follows a similar approach and
proposes a solidity-like language with the addition of state and state
transitions as first-class constructs in the programming language.
These proposals all aim to make the contracts more predictable.
\emph{Simplicity}~\cite{OConnor2017-simplicity} exhibits larger
expressiveness yet allowing easy static analysis compared
to EVM code. Static analysis provides useful upper bound computation estimation
on the transactions, thus giving the peers more predictable views on the
transaction execution. Simplicity also features self-contained transactions
that exclude the global state in the contract execution.

Notice that such novel programming languages do make it simpler for developers to write correct
contracts. However, wide-scale deployment of new programming models would
require rewriting of all legacy software, which requires significant
development effort.

%% file: sections/conclusion.tex

Re-entrancy attacks exploit inconsistent internal state of smart contracts
during unsafe re-entrancy, allowing an attacker in the worst case to drain all
available assets from a smart contract. So far, it was believed that advanced
offline analysis tools can accurately detect these vulnerabilities. However, as
we show, these tools can only detect basic re-entrancy attacks and fail to
accurately detect new re-entrancy attack patterns, such as cross-function,
delegated and create-based re-entrancy.
Furthermore, it remains open how to protect existing contracts as smart
contract code is supposed to be immutable and contract creators are anonymous,
which impedes responsible disclosure and deployment of patched contract.
To address the particular ecosystem of smart contracts, we introduce a novel
run-time smart contract security solution, called \toolname, which exploits
dynamic taint tracking to monitor data-flows during smart contract execution to
automatically detect and prevent inconsistent state and thereby effectively
prevent basic and advanced re-entrancy attacks without requiring any semantic
knowledge of the contract. By running \toolname\ on almost 80 million Ethereum
transactions involving 93,942 contracts, we show that \toolname\ can prevent
re-entrancy attacks in existing contracts with negligible overhead.
\toolname\ is designed to run in enforcement mode, protecting existing
contracts, when \toolname\ is integrated into the blockchain ecosystem.
However, \toolname\ can be particularly relevant for smart contract developers
in order to identify attacks against their contracts and patch them
accordingly. Namely, \toolname\ can also be executed locally by contract
developers that are interested in ensuring the security of their deployed
contracts.
Lastly, we are the first in presenting and analyzing false positive cases when
searching for re-entrancy vulnerabilities. We reveal root causes of false
positive issues in existing approaches and give concrete advice to smart
contract developers to avoid suspicious patterns during development.